\documentclass[]{aastex631}
\accepted{July 14, 2023}

\submitjournal{ApJ}

\usepackage{xcolor}
\usepackage{physics}
\usepackage[autostyle]{csquotes}
\usepackage{ gensymb }
\usepackage[T1]{fontenc}

\usepackage[shortcuts]{extdash}
\newcommand\LST{LST\=/1}

\begin{document}

\title{Observations of the Crab Nebula and Pulsar with the Large-Sized Telescope Prototype of the Cherenkov Telescope Array}

\author{H.~Abe}
\affiliation{Institute for Cosmic Ray Research, University of Tokyo, 5-1-5, Kashiwa-no-ha, Kashiwa, Chiba 277-8582, Japan}
\author{K.~Abe}
\affiliation{Department of Physics, Tokai University, 4-1-1, Kita-Kaname, Hiratsuka, Kanagawa 259-1292, Japan}
\author{S.~Abe}
\affiliation{Institute for Cosmic Ray Research, University of Tokyo, 5-1-5, Kashiwa-no-ha, Kashiwa, Chiba 277-8582, Japan}
\author{A.~Aguasca-Cabot}
\affiliation{Departament de Física Quàntica i Astrofísica, Institut de Ciències del Cosmos, Universitat de Barcelona, IEEC-UB, Martí i Franquès, 1, 08028, Barcelona, Spain}
\author[0000-0002-3777-6182]{I.~Agudo}
\affiliation{Instituto de Astrofísica de Andalucía-CSIC, Glorieta de la Astronomía s/n, 18008, Granada, Spain}
\author{N.~Alvarez~Crespo}
\affiliation{IPARCOS-UCM, Instituto de Física de Partículas y del Cosmos, and EMFTEL Department, Universidad Complutense de Madrid, E-28040 Madrid, Spain}
\author[0000-0002-5037-9034]{L.~A.~Antonelli}
\affiliation{INAF - Osservatorio Astronomico di Roma, Via di Frascati 33, 00040, Monteporzio Catone, Italy}
\author[0000-0002-8412-3846 ]{C.~Aramo}
\affiliation{INFN Sezione di Napoli, Via Cintia, ed. G, 80126 Napoli, Italy}
\author{A.~Arbet-Engels}
\affiliation{Max-Planck-Institut für Physik, Föhringer Ring 6, 80805 München, Germany}
\author{C.~Arcaro}
\affiliation{INFN Sezione di Padova and Università degli Studi di Padova, Via Marzolo 8, 35131 Padova, Italy}
% ...



\begin{abstract}
CTA (Cherenkov Telescope Array) is the next-generation ground-based observatory for gamma-ray astronomy at very-high energies. The Large-Sized Telescope prototype (\LST{}) is located at the CTA-North site, on the Canary Island of La Palma. LSTs are designed to provide optimal performance in the lowest part of the energy range covered by CTA, down to $\simeq 20$ GeV. \LST{} started performing astronomical observations in November 2019, during its commissioning phase, and it has been taking data since then. We present the first \LST{} observations of the Crab Nebula, the standard candle of very-high energy gamma-ray astronomy, and use them, together with simulations, to assess the performance of the telescope. \LST{} has reached the expected performance during its commissioning period - only a minor adjustment of the preexisting simulations was needed to match the telescope behavior. The energy threshold at trigger level is around 20 GeV, rising to $\simeq 30$ GeV after data analysis. Performance parameters depend strongly on energy, and on the strength of the gamma-ray selection cuts in the analysis: angular resolution ranges from 0.12 to 0.40 degrees, and energy resolution from 15 to 50\%. Flux sensitivity is around 1.1\% of the Crab Nebula flux above 250 GeV for a 50-h observation (12\% for 30 minutes). The spectral energy distribution (in the 0.03 - 30 TeV range) and the light curve obtained for the Crab Nebula agree with previous measurements, considering statistical and systematic uncertainties. A clear periodic signal is also detected from the pulsar at the center of the Nebula.

\end{abstract}


\section{Introduction} 

 Astronomical observations across the electromagnetic spectrum span more than twenty decades in photon energy, requiring the use of a wide range of instrumental techniques. At the high-energy end of the spectrum, in the so-called very-high-energy (VHE) gamma-ray band and above (E$_\gamma \gtrsim$ 50 GeV), photons are scarce, and the performance of space-borne facilities is limited by their modest photon collection areas. Ground-based instruments \citep{2022Galax..10...21S}, on the other hand, can achieve much larger effective areas by exploiting the effects of the absorption in the atmosphere of a VHE photon: the development of an extensive air shower (EAS) of secondary particles. Imaging Atmospheric Cherenkov Telescopes (IACTs) collect the Cherenkov light produced by the ultra-relativistic EAS particles (mostly electrons and positrons), and form an image of the shower, from which the direction and energy of the primary photon can be estimated.

 The Crab Nebula was the first source detected in the VHE sky \citep{Whipple_Crab}, now featuring around 250 known objects \citep{2022Galax..10...21S}. The Crab Pulsar Wind Nebula (PWN) is the remnant of a Supernova explosion observed in 1054 A.D. It is one of the most studied objects in the sky \citep{2008ARA&A..46..127H, 2014RPPh...77f6901B} and its spectrum extends from radio up to PeV gamma rays \citep{LHAASO_Crab}. Flaring or flickering behavior has been reported below a few GeV \citep{doi:10.1126/science.1199705}, but in the VHE band it remains a steady source \citep{HESS_flares, VERITAS_flares}, and is currently the standard candle used by different instruments \citep{CrabMAGIC, HESS_Crab}. 

The Cherenkov Telescope Array\footnote{\url{https://www.cta-observatory.org/}} (CTA) is the next-generation ground-based observatory for VHE gamma-ray astronomy. CTA will consist of two arrays of IACTs of different sizes deployed on two  sites: one in the Northern Hemisphere, in the {\it Roque de los Muchachos} observatory (ORM) in the Canary island of La Palma, and another one in the Southern Hemisphere, in the desert of Atacama in Chile. The largest among the CTA telescopes are the Large-Sized Telescopes\footnote{\url{https://www.lst1.iac.es/}} (LSTs, \citealp{Cortina:2019juz}), four of which will be part of the CTA-North array. LSTs are equipped with 23 m diameter mirror dishes, which enable them to detect the faint Cherenkov flashes from showers initiated by gamma rays down to $\simeq 20$ GeV. The lightweight LST structure is designed to allow fast slewing, and hence facilitate follow-up observations of transients. \LST{}, the prototype of the LST, was inaugurated at the ORM in October 2018 and has been taking science data since November 2019.

This paper presents the observations of the Crab Nebula performed with \LST{} between November 2020 and March 2022, while the telescope was still in its commissioning phase. These observations are used to characterize the performance of the instrument, and to validate the Monte Carlo (MC) simulations needed for the data analysis. Section \ref{sec:mc} describes the MC simulations. The data analysis pipeline is summarized in section \ref{sec:pipeline}, which also presents some key \LST{} performance parameters (energy and angular resolution) as determined from the MC. Section \ref{sec:data_sample} introduces the Crab Nebula data sample, which in section \ref{sec:data_mc_comparison} is used to verify the validity of the simulations. Section \ref{sec:results} presents the results on the Crab Nebula spectrum and light curve, and an estimate of the \LST{} sensitivity, and finally we provide some conclusions in Section \ref{sec:conclusions}

\section{Monte Carlo simulations \label{sec:mc}}
An IACT uses the atmosphere as a key element in the detection process, hence there is obviously no way to characterize its end-to-end response in a controlled set-up (unlike space-borne gamma-ray telescopes, which can be beam-tested in the lab prior to launch). The analysis of IACT images to reconstruct the properties of the shower primary (particle identity, energy and direction) relies heavily on detailed MC simulations both of the shower development and of the telescope. The simulated events, for which we know the true attributes of the primary particle, will allow us to train the event reconstruction algorithms to be used on the real data. The same algorithms are also applied to an independent MC sample (\enquote{test} sample), from which the instrument response functions (IRFs) can be derived.

\subsection{General Description}

We simulate the shower development in the atmosphere, and the Cherenkov light emission and propagation, using \texttt{CORSIKA} v7.7100 \citep{1998cmcc.book.....H}. For the telescope simulation we use \texttt{sim\_telarray} v2020-06-28 \citep{2008APh....30..149B}. The telescope simulation includes the reflection of the Cherenkov light on the mirror dish, its passage through the camera entrance window, and its detection on the focal plane equipped with light concentrators and photomultiplier tubes (PMTs). All the stages are simulated taking into account the lab-measured performance of the different telescope elements as a function of the photon wavelength and, when relevant, its incident direction. The camera trigger system, the electronic chain for the signal processing and the digitization of the analog signals are also simulated in detail to obtain digital waveforms (two per pixel, from a high- and a low-gain branch) completely equivalent to those present in the real \LST{} data after the correction of low-level features of the data acquisition electronics (see section \ref{sec:pulse_integration}). From this point onward, the same analysis pipeline, described in section \ref{sec:pipeline}, can be applied to simulated and observed data.

 To train the event reconstruction algorithms (training MC sample), we use simulations of two types of primary particles: gamma rays and protons. To evaluate the response of the telescope (test MC sample), we only simulate gamma rays. For the analysis of gamma-ray observations it is not strictly necessary to obtain from MC simulations the response of the telescope to background events (showers initiated by electrons, protons, and heavier cosmic-ray nuclei). The precise simulation of the background of an IACT is challenging, due to the uncertainties in the hadronic interaction models \citep[see e.g.][]{Ohishi:2021hfu} and in the cosmic-ray composition, but, as we will see, a reliable estimation of the background event rates recorded from the direction of the source of interest can be obtained using a control (off-source) sky region, i.e. using simple aperture photometry. Finally, we also produce simulations of single muons, which are helpful to evaluate the overall optical efficiency of the telescope \citep{muons_cta}. The main simulation parameters for the production of each type of primary are presented below.

\subsection{MC Generation Limits}
\subsubsection{Gamma Rays}
Gamma rays are simulated with a differential energy spectrum dN/dE $\propto\,$E$^{-2}$. For vertical incidence the energy range was set to 5 GeV - 50 TeV. In order to account for the variation of the telescope energy threshold with the zenith distance (ZD), these energy limits are then modified as E$_{\text{min, max}} \propto \cos^{-2.5}(\text{ZD})$, with a maximum value of 200 TeV. 

In the training sample the directions of the gamma rays are distributed isotropically around the telescope pointing up to an offset angle of 2.5$^\circ$, such that the event reconstruction algorithms can be trained (see section \ref{sec:rf}) to perform well for sources anywhere in the field of view. In order to speed up the simulations, and considering the low global trigger efficiency, each generated shower is used 10 times, placing the telescope at 10 different random locations around the shower axis, out to a maximum impact parameter of 900 m for ZD = 0.  This value is scaled as $\cos^{-0.5}$(ZD) for larger zenith distances.

The gamma-ray test sample is produced in a similar way, although for the analysis of standard observations of point-like sources, in which the telescope is pointed at a sky position 0.4$^\circ$ away from the source (see section \ref{sec:data_sample}), the gamma-ray directions are all generated with that precise offset from the telescope pointing (in random orientations). The small offset angle allows reducing the maximum impact to 700 m for vertical incidence, which is then scaled as $\cos^{-1}$(ZD).

From the dependence of the air mass as $\cos^{-1}$(ZD) in the plane-parallel approximation, one would naively expect (for electromagnetic showers and in absence of atmospheric absorption) the same evolution of the distance to the shower maximum, and hence of the radius of the Cherenkov light pool. The photon density reaching the mirror dish would change as $\cos^2$(ZD) under the same assumptions. This would suggest using $\cos^{-2}$(ZD) and $\cos^{-1}$(ZD) scalings for the energy generation limits and the maximum impact parameter respectively. These simple zenith-scaling laws were re-evaluated empirically using an earlier simulation library with protons and gamma rays simulated at zenith distances of 20, 40 and 60 degrees. The power index for the energy generation limits was corrected to -2.5 (a faster increase with zenith) to keep the fraction of triggers (and hence the computational cost of producing triggered events) closer to the one at zenith. As for the maximum impact parameter, the $\cos^{-1}$(ZD) dependence worked well for the gamma-ray test sample, whereas $\cos^{-0.5}$(ZD) (a slower increase) was chosen for the diffuse gamma simulations (training sample), to moderate the drop in the production efficiency.

\subsubsection{Protons}
Protons for the training sample are generated following the same scaling of maximum impact parameter and energy range as for diffuse gamma rays. The energy range for vertical incidence is 10 GeV - 100 TeV (taking into account their lower Cherenkov light yield compared to gamma rays of the same energy), and the maximum energy is also capped at 200 TeV. The maximum impact parameter is 1500 m for vertical showers, and the number of shower re-uses is also set at 10. Directions are isotropically distributed within $8 \degree$ of the telescope pointing, with that maximum offset changing as $\cos^{0.5}(\text{ZD})$ - once again, a dependence obtained empirically to maintain the production efficiency similar to the one at zenith. For hadronic showers, as compared to electromagnetic ones, it is harder to estimate zenith-dependent optimal production ranges from simple arguments, due to the production of electromagnetic sub-showers and muons at large angles relative to the primary particle.

The MC production ranges and zenith-scaling laws above may be revised in the near future to further reduce the computational cost of the Monte Carlo generation. Note that we do not need a \enquote{complete} MC proton training sample, in the sense of containing all (or a very large fraction of all) the protons that would trigger the telescope. Its only purpose is to be used in the training of the particle classification algorithm together with the gamma rays, and this can be achieved as long as the simulated protons produce a sufficiently representative sample of the hadronic shower images that the real background (protons and other nuclei) will generate. On the other hand, we cannot exclude that a {\it more complete} training sample (more computationally expensive) could bring some improvement to the performance we present in this paper.

\subsubsection{Muons}
Single muons are simulated with a power-law differential energy spectrum with index -2.0, between 8.4 GeV and 1 TeV. The maximum impact parameter is 9.8 m, corresponding to 80\% of the mirror radius, since we are interested in muons which produce large rings on the camera. The maximum off-axis angle is $0.9 \degree$, which means the rings would be fully contained on the camera. Muons are simulated with vertical incidence. The amount of Cherenkov light produced by a single muon and collected by the telescope correlates very well with its impact parameter, direction of incidence and Cherenkov emission angle (which in turn depends on its energy). All these parameters can be reconstructed from the ring-shaped image formed on the camera, which makes isolated muons a perfect tool for the calibration of the total light throughput of the telescope \citep{muons_cta}. This is achieved through the comparison of the actual observations with muon simulations performed with different global optical efficiency of the telescope. For the present work we simulated 10$^4$ muons per optical efficiency, which we vary in steps of 10\%.

\subsection{MC Pointing Grids}
Simulations of gamma rays and protons were performed in a wide range of telescope pointing directions, up to $\simeq 70^\circ$ zenith distance. These directions were chosen in two different grids of pointings, one for the training MC and one for the test MC.

\begin{figure}
    \centering
    \includegraphics[width=0.48\textwidth]{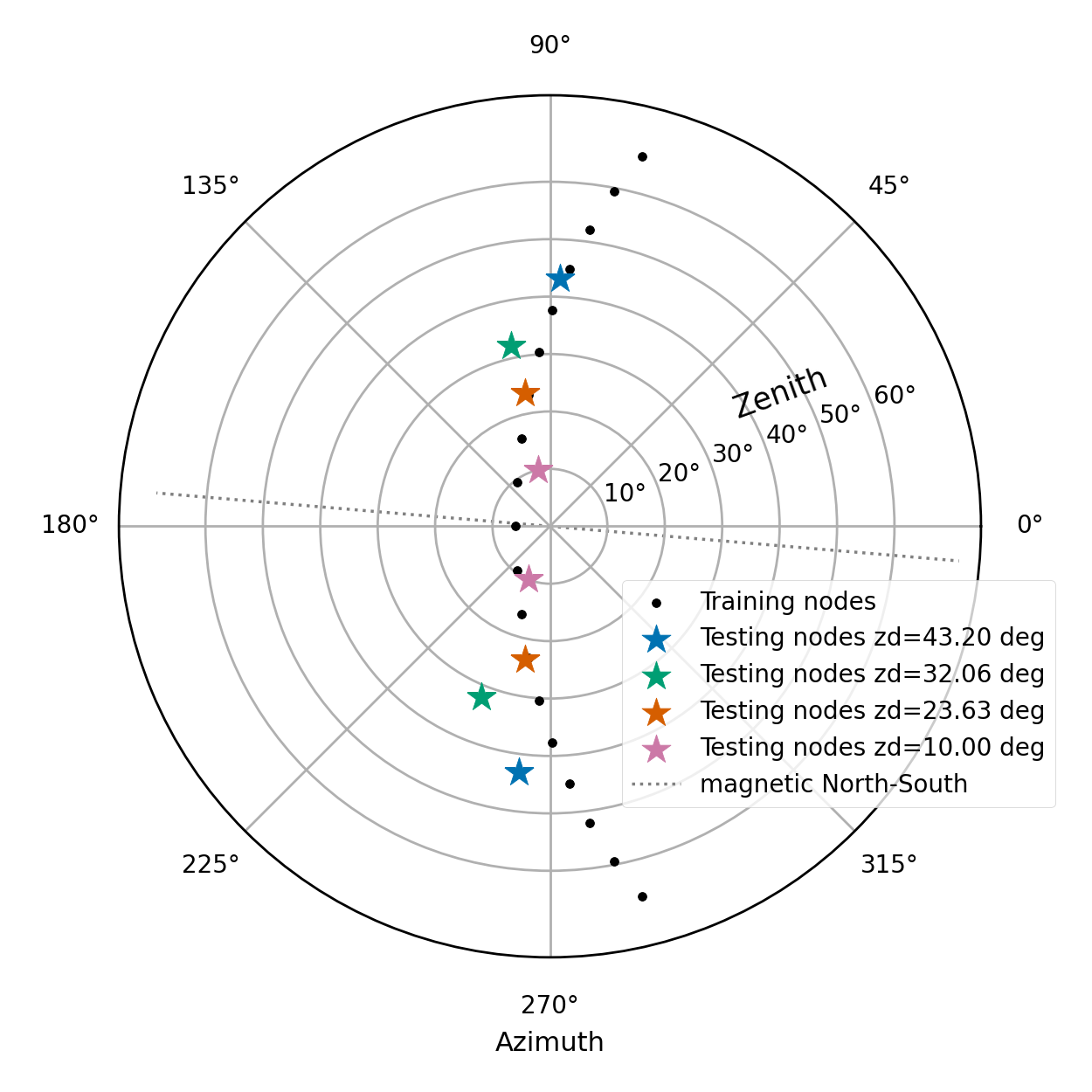}
    \caption{Pointing directions of the nodes. The training nodes are used to train the shower reconstruction algorithms. The testing nodes are used to compute the IRFs from point source gamma-ray simulations. The nodes that have the same altitude and symmetrical azimuths with respect to the magnetic North are combined (by color here) to compute the IRFs in Fig. \ref{fig:irfs_srcindep}.}
    \label{fig:pointing_nodes}
\end{figure}

\subsubsection{Training MC} 
The training MC is simulated using pointings along \enquote{declination lines}, i.e. following the trajectory in horizontal (Alt-Az) coordinates that all sources at a given declination follow as viewed from the \LST{} site. We simulated 15 different declination lines (from -$29\degree$ to +67$^\circ$, distributed in steps of cos(ZD$_\text{min}$), where ZD$_\text{min}$ is the zenith angle at culmination. In each line around twenty different pointings equally spaced in hour angle are defined, with the limits being determined as: ZD < 70$^\circ$ and $\pm$ 6 hours around culmination. For the analysis of the Crab Nebula observations presented here we used the closest declination line, which corresponds to culmination at ZD = 6$^\circ$ (i.e. $\delta = 22.76^\circ$), and has a total of 19 pointings, see Fig. \ref{fig:pointing_nodes}. The rationale behind this grid design is to train the algorithms with the MC from all the directions along the chosen declination line, including the telescope (Alt-Az) pointing coordinates among the parameters made available as input to the reconstruction (see section \ref{sec:rf}). In this way, the algorithms can automatically {\it learn} how image parameters vary with the incidence direction of the shower primary (mainly through the variations of airmass with zenith distance). In principle, one could also train the algorithms with \enquote{all-sky} simulations, i.e. the full grid of pointings, and obtain algorithms that would be able to reconstruct events from any direction, regardless of the sky coordinates of the observed source. However, this would result in algorithms with significantly higher memory needs (for our declination-wise approach, the size of the four random forests is already 18.1 Gbyte, that have to be kept in memory during the analysis), and no improvement in performance, since the analysis of any given source would only make use of the pointings along the source path.

\subsubsection{Test MC}
The test MC is simulated in a grid of zenith and azimuth values. For a single IACT, the relevant direction-dependent quantities which affect the performance are the airmass (which increases like 1 / cos(ZD) in the planar atmosphere approximation) and the component of the geomagnetic field orthogonal to the shower axis (which is the relevant one affecting the shower shape, via the Lorentz force acting on the secondary charged particles). We therefore produced a regular triangular grid in $\cos(\text{ZD})$ and B$_\perp$/B. Eventually we will use this test MC grid to calculate the IRFs in each node, and then interpolate them to obtain the IRFs for any arbitrary telescope pointing, but this procedure is still in development; for the present analysis we will simply use for each data run the IRFs obtained with the closest test MC grid node. Two nodes at ZD=10$^\circ$ and two nodes at ZD=32$^\circ$ from the regular grid have been used in this work for the analysis of observations within 35$^\circ$ of zenith. Besides, we also used MC generated for two additional pointings at ZD=23.6$^\circ$ (along the Crab path, on both sides of culmination) to improve the precision of the IRFs, considering that we are using no interpolation to the actual telescope orientation. Fig. \ref{fig:pointing_nodes} shows all the pointing directions in our test MC sample.

\subsection{Adjustments of the MC Telescope Simulation} 

The MC simulation of the Cherenkov light detection is initially based on lab measurements of the individual response of the different telescope elements (e.g. the reflectivity of the mirrors, or the photon detection efficiency of the PMTs). Typically, the simulation parameters need some tuning in order to match the overall performance of the telescope once deployed on site, where it is subject to various sources of degradation (dirt on mirrors or on the camera window, varying atmospheric conditions, different levels of background light, temporarily misaligned mirror tiles...). Below we describe the adjustments performed on the \LST{} MC simulation for the analysis of the data taken during the commissioning period.

\subsubsection{Optical Efficiency}
\begin{figure*}
\begin{center}
\includegraphics[width=0.49\textwidth]{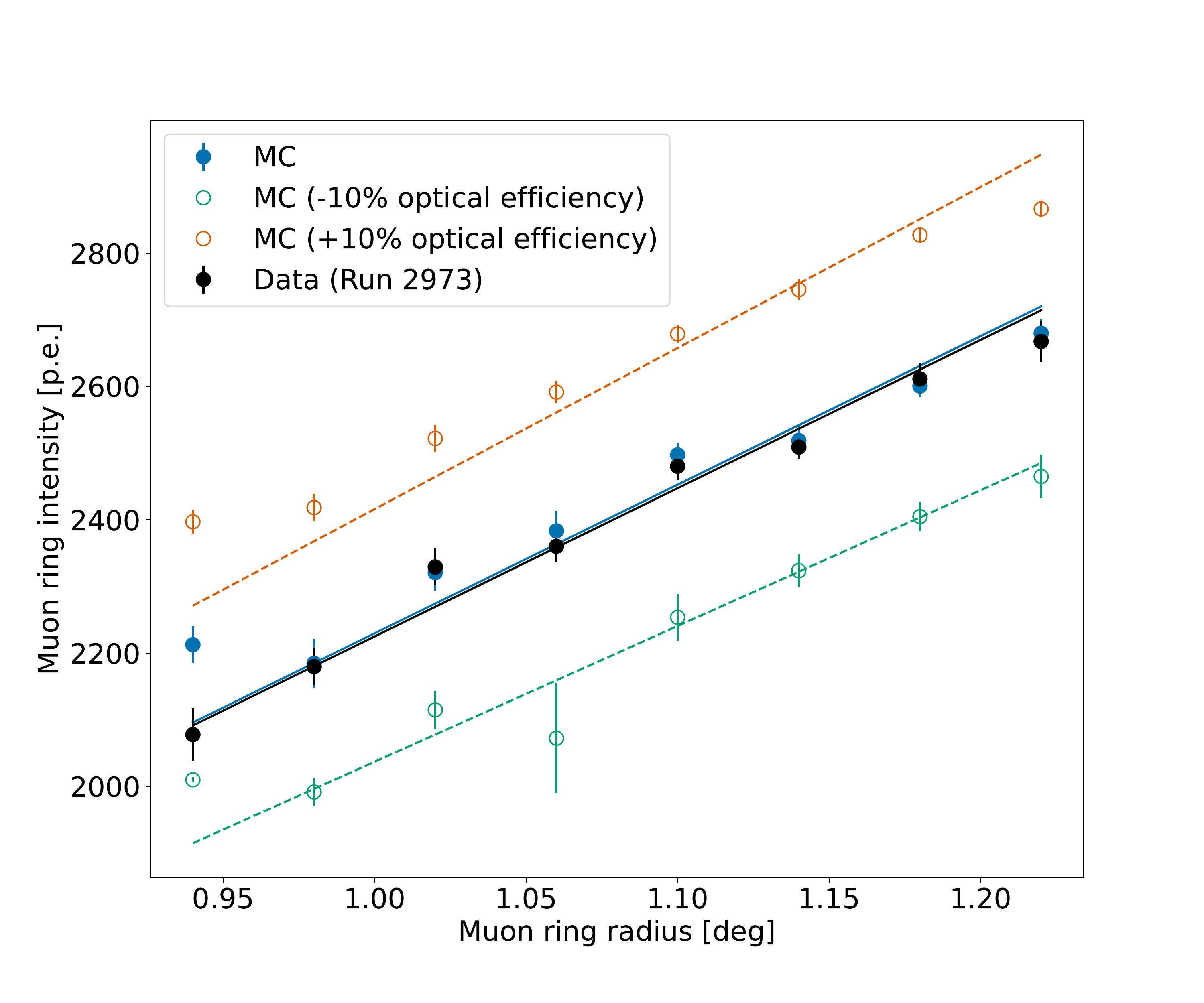}
\includegraphics[width=0.49\textwidth]{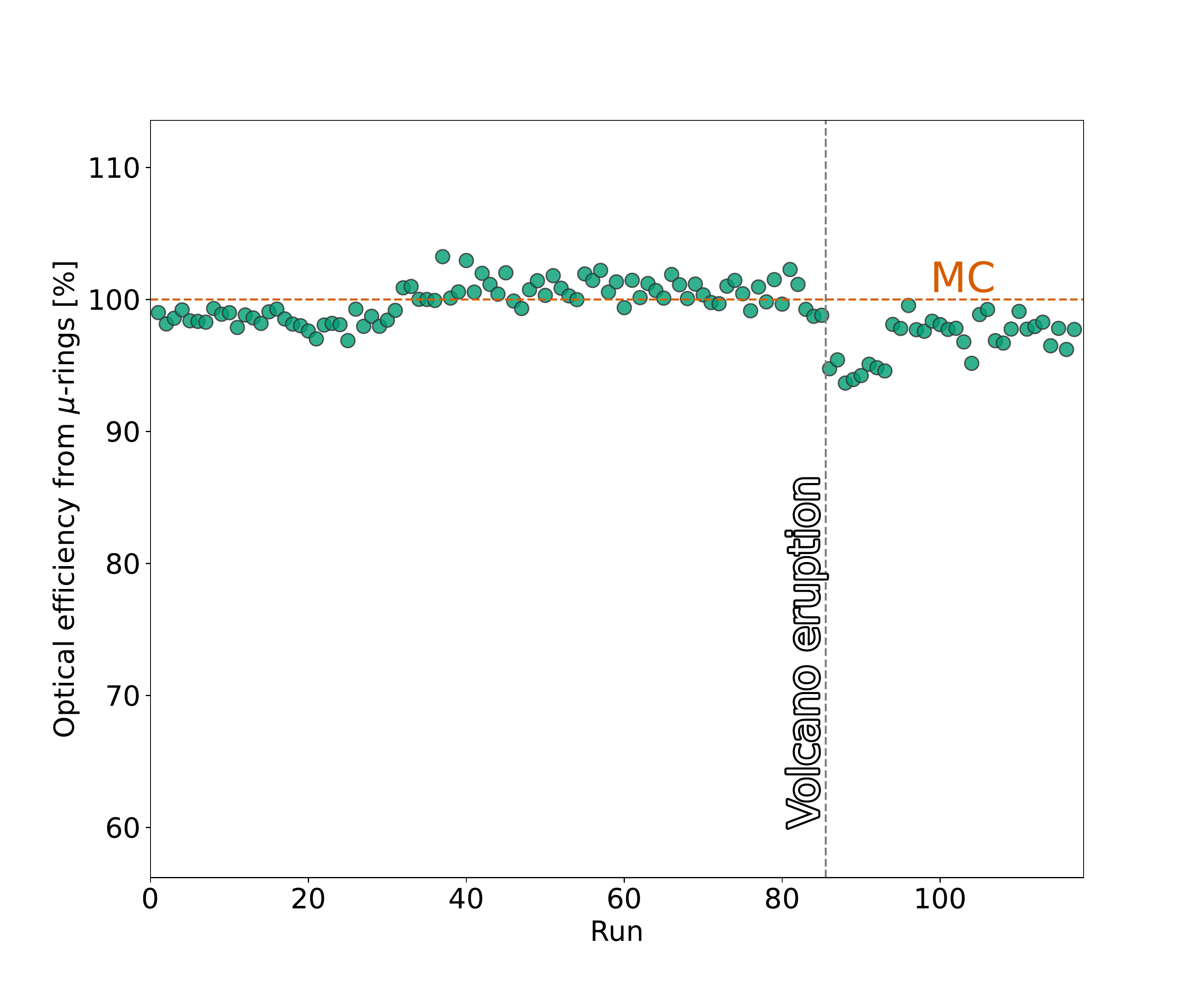}
\vspace{-0.3cm}
\caption{Left panel: total light in muon rings detected by \LST{} vs. ring radius. Data from one run is compared to three different simulations, with telescope efficiency changing in steps of 10$\%$. Right panel: run-wise \LST{} optical efficiency derived using muon ring intensity, for the sample of good-quality Crab Nebula observations analyzed in this work. The vertical dashed line separates the runs taken before and after the volcanic eruption in La Palma (which took place between 19/09 and 13/12 2021). The observed variations are mostly driven by changes in the number of correctly aligned mirrors, and in their reflectivity. For example, the early post-eruption data indicated a $\simeq 5\%$ lower efficiency, which partially recovered after some rainfall in February 2022, which presumably removed dirt from the dish surface.}
\label{fig:muon}
\end{center}
\end{figure*}

The overall optical efficiency of an IACT can be affected by mirror degradation or dust deposit on its surface, that would reduce its reflectivity. The optical efficiency in the MC is tuned to the real optical efficiency of the telescope obtained through the analysis of muon rings \citep{muons_cta}. We monitor this optical efficiency in the daily data checks to spot possible problems such as misaligned mirrors, or loss of reflectivity. The method works as follows: for muon rings (unlike for shower images, see section \ref{sec:pulse_integration}) we obtain the pixel-wise charges with an unbiased pulse integrator, i.e. one with a common integration time window for all pixels, defined around the camera-averaged peak time of the muon light pulse. Pixels not illuminated by the muon, containing only noise, will on average have zero charge - the unbiased integrator will not pick preferentially positive fluctuations of the Night Sky Background light. Then a fit is performed to obtain the geometrical ring parameters (center and radius R). The total light (or muon ring intensity) is obtained by adding up the charges in all pixels whose centers are at a distance between 0.75\,R and 1.25\,R from the ring center. This integration range is large enough to contain all of the muon light, considering the optical point-spread function (PSF) of the telescope, and the possible inaccuracies of the ring fit. The muon ring intensity is expected to be proportional to the muon ring radius (which is the Cherenkov angle of the muon), see Eq. 21 of \cite{muons_cta}. The left panel of Fig. \ref{fig:muon} shows this approximate behavior for the rings recorded in one \LST{} data run, compared to MC simulations with three different efficiencies.
 
 The optical efficiency of the telescope may fluctuate on a nightly basis at the few percent level, as can be seen in the right panel of Fig. \ref{fig:muon}, which shows its evolution as measured from well-contained muon rings in the sample of 117 good-quality Crab observation runs (of a typical duration of 20 minutes) spanning $\simeq 1.5$ years, that will be described in section \ref{sec:data_sample}. Given the small variations of the muon light yield, for the analysis presented in this paper no run-wise tuning has been introduced; a single optical efficiency (indicated as ``100\%'' in Fig. \ref{fig:muon}) was used in the MC simulations used for the analysis of the whole data sample.

\subsubsection{Optical Point Spread Function}
The optical Point Spread Function (PSF) of the \LST{} mirror dish is measured by a dedicated CCD camera located at its center, which records images of stars focused on a special removable screen located between the focal plane and the camera entrance window. The simulation can be tuned to match the real PSF through two parameters which determine the spread in the orientations of the individual mirror tiles relative to their ideal alignment. The quality of the PSF of the telescope may fluctuate due to e.g. sporadic problems in the active mirror control system, but in the period considered in this work no significant variations were recorded which would require separate MCs with different adjustments. Therefore, for the analysis presented in this paper, the same PSF was used in the MC simulations of the telescope. With the selected settings, 80\% of the PSF is contained in a circle with a diameter of $\simeq 0.06 \degree$, i.e. well within one pixel ($\diameter=0.1\degree$).

\subsubsection{Night Sky Background}
In the \LST{} simulations with \texttt{sim\_telarray} the assumed level of Night Sky Background light (NSB) is that of a \enquote{dark} sky field (typical of observations away from the Galactic disk and the zodiacal plane). It results in an average photoelectron (p.e.) rate of 193 MHz per pixel.

\LST{} observations, on the other hand, are performed in a wide range of NSB conditions (even with the Moon above the horizon). Instead of re-running the telescope simulation for different levels of NSB, which would be computationally expensive, we adopt the approach known as \enquote{noise padding}, which consists in adding some random Poissonian noise during the analysis of the MC events (right before the image cleaning - see section \ref{sec:pipeline}), to match the average NSB level in the data. The level of NSB in a given field of view is measured using the interleaved pedestal events acquired during observations at a rate of 100 Hz (these are events containing only noise). Since all the data analyzed in this work were recorded in dark night conditions (with the Moon below the horizon), and pointing towards the same sky region, the whole sample can be processed with a single NSB tuning of the MC (which corresponds to $\simeq$ 70\% larger NSB p.e. rate than the default in the simulation).

\section{Pipeline description \label{sec:pipeline}}

\texttt{cta-lstchain} \citep{lstchain} is the data analysis pipeline used to process the \LST{} data. It is heavily based on \texttt{ctapipe} \citep{ctapipe, Noethe:20219K}, the prototype low-level analysis framework of CTA. \texttt{cta-lstchain} uses the \texttt{ctapipe\_io\_lst}\footnote{\url{https://github.com/cta-observatory/ctapipe_io_lst}} plug-in to read in the raw data files, and to apply a first calibration to the camera signals. \texttt{cta-lstchain} performs the event reconstruction as described below, and outputs files containing lists of gamma-ray candidate events (including event-wise reconstructed parameters like energy and incoming direction) writing out DL3 data in the format specified by the Data Formats for Gamma-Ray Astronomy (GADF) \citep{deil_GADF, 2019A&A...625A..10N}. These files can be further processed with Gammapy \citep{Gammapy_proceeding, gammapy}, the official high-level data analysis framework of CTA, to obtain final analysis products like energy spectra or light curves. The IRFs accompanying the selected gamma-ray candidate events in the DL3 files are produced using \texttt{pyirf} \citep{pyirf}. Most of the data reduction was carried out in an on-site computing infrastructure at La Palma using \texttt{lstosa} \citep{lstosa_ADASS, lstosa_code}, a library which automatizes the processing of the \LST{} data through the \texttt{cta-lstchain} pipeline.

In the rest of this section we provide details on the different stages of the event reconstruction carried out by \texttt{cta-lstchain}.

\subsection{Pixel-wise Charge Integration \label{sec:pulse_integration}}  
For every triggered event, the raw data recorded by the \LST{} camera consists of 12-bit digitized waveforms. Two such waveforms (corresponding to the high- and low-gain branches) are stored per pixel, each having 40 samples taken at 1 ns intervals. The raw waveforms are calibrated as described in \cite{Kobayashi:2021jc}. In most cases the calibrated high-gain waveform is the one used for further analysis; only when it has one or more raw samples above 3500 counts is the high gain discarded and the low gain is used instead (the transition occurs at a charge of around 200 p.e. per pixel). Each waveform is then integrated in an 8-sample range (1 sample $\simeq$ 1 ns), using the so-called LocalPeakWindowSum algorithm, in the range $[s_{max}-3, s_{max}+4]$, where $s_{max}$ is the sample with the highest value in the calibrated waveform. The integrated charge is converted to photoelectrons using conversion factors determined (using the so-called F-factor method) in the analysis of dedicated calibration runs \citep{Kobayashi:2021jc}. The conversion factors include a correction for the (average) pulse tails beyond the integration window, which typically contains around 84$\%$ of the total charge. Besides the integrated charge, for each pixel we calculate a signal arrival time using a simple charge-weighted average of the sample times.

\subsection{Image Cleaning and Parametrization}  

The vast majority of pixels in each camera event contain only noise, fluctuations of the night sky background light. A procedure called \enquote{image cleaning} is applied to each event to select the pixels which contain a significant amount of Cherenkov light from the shower. Two charge thresholds are defined, with default values 8 and 4 p.e. - known as \enquote{picture} and \enquote{boundary} thresholds respectively \citep{LESSARD20011}. A first pixel selection is performed applying the following algorithm:
\begin{itemize}
    \item[i)] increase the picture threshold for pixels which have a large level of noise: we will replace the default value by $\langle Q_{\text{ped}} \rangle + 2.5 \cdot \sigma_{\text{Qped}}$, if that value happens to be larger than the default 8 p.e. The quantities $\langle Q_{\text{ped}} \rangle$ and $\sigma_{\text{Qped}}$ are the mean and the standard deviation of the reconstructed charge of that specific pixel, as computed from interleaved pedestal events.
    \item[ii)] select pixels with charge above the picture threshold, and with at least two neighboring pixels above the picture threshold. The pixels fulfilling this condition form the \enquote{core} of the image.
    \item[iii)] select pixels with charge above the boundary threshold and at least one core neighbor.
\end{itemize}

{\setlength{\parindent}{0cm} We then apply the following additional conditions to keep a pixel as part of the final image:}

\begin{itemize}
    \item[iv)] to have at least one neighbor (among the pre-selected set of pixels described above) with signal arrival time within 2 ns of its own arrival time.
    \item[v)] to have a charge above $0.03 \cdot Q_{peak}$, where $Q_{peak}$ is the average charge of the three brightest pixels in the image.
\end{itemize}

A convenient way of assessing the effectiveness of the image cleaning algorithm is to check the probability that a pedestal event survives the procedure (i.e. it has a non-zero number of surviving pixels). That probability will be the same with which spurious islands of noise-only events survive the cleaning in the analysis of genuine shower images. Such noise-only islands have the potential of spoiling the shower reconstruction. Conditions (i) to (iv) are sufficient, for observations with \LST{} in dark conditions (with the Moon below the horizon), to have a fraction < O($10^{-3}$) of cleaning-surviving pedestal events.

The step (i) is necessary in order to ensure that bright stars in the field of view (which increase in the illuminated pixels the rate of signals above the default picture threshold) do not produce a significant number of spurious islands. The values of $\langle Q_{\text{ped}} \rangle$ and $\sigma_{\text{Qped}}$ are calculated every few seconds using interleaved pedestal events, to take into account the rotation of the star field during an observation. This \enquote{anti-star} condition will increase the effective cleaning threshold in the pixels around stars, and in turn produce a deficit of dim images in those areas, i.e. a non-uniformity in the response of the camera at low energies. In order to limit this effect, it is important that the {\it default} picture threshold is high enough so that only a small part of the camera pixels get their picture thresholds modified in step (i). In this way we will still be able to use for the subsequent analysis a simplified MC simulation with a uniform NSB level across the field of view (and hence uniform camera response). For the default picture threshold of 8 p.e., and the sample analyzed in this paper, only 4.7$\%$ of the camera pixels (on average) end up with increased values due to the effect of stars (and only 1.2\% have a value above 10 p.e.).

Conditions (ii), (iii) and (iv) are intended to select groups of neighboring pixels with significant signals arriving close in time, as expected for showers (and reject noise fluctuations, which are not correlated in different pixels). Condition (v) is introduced to solve a problem present in some of the \LST{} data taken in the commissioning phase: occasionally some mirror tiles were purposefully misaligned (making them point $\simeq2^\circ$ away from the main spot), whenever they could not be accurately adjusted. In this way they did not deteriorate the optical PSF of the dish. However, the range of the mirror orientation mechanism is not enough to make those mirrors point always outside of the camera limits, which means that they can produce dimmer \enquote{duplicate} images of very bright showers in different positions on the camera. By requiring a minimum pixel charge of 3$\%$ of the peak charge in the event, we removed all such fake images present in the data. This cut also removes some pixels in the actual tails of (very bright) images, but we verified via MC simulation that this had no negative impact on the performance of the standard event reconstruction described in this paper (including the gamma / hadron separation capabilities).

The cleaned images are then parametrized by a modified version of the Hillas parameters, first described in \cite{1985ICRC....3..445H}, and higher-order moments. The complete list of parameters and their description is presented in the Appendix. These parameters are fed to machine learning algorithms (random forest, \citealt{breiman}) for the subsequent steps in the event reconstruction.

\begin{figure}
    \centering
    \includegraphics[width=0.49\textwidth]{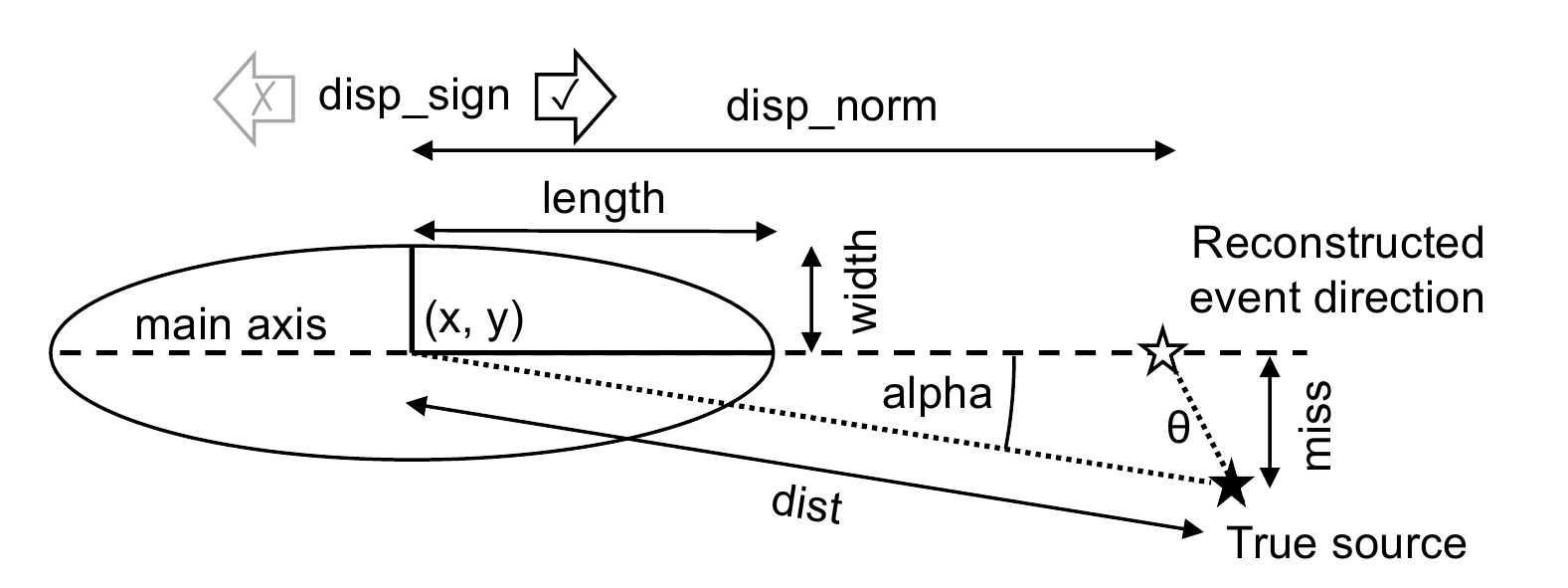}
    \caption{Definition of basic image parameters}
    \label{fig:image:sketch}
\end{figure}

\subsection{Random Forest Training \label{sec:rf}}

Image parameters are used to train random forests using \texttt{scikit-learn} \citep{scikit-learn} to reconstruct the desired physics parameters: the direction and energy of the primary, and a {\it score} (which we call {\it gammaness}) indicating how likely it is that the primary is a gamma ray, rather than a proton or other cosmic-ray particle. The hyperparameters of the models are presented in the Appendix for complete reproducibility.

\subsubsection{Direction Reconstruction}
\begin{figure}
    \centering
    \includegraphics[width=1.0\linewidth,trim={3cm 4cm 3cm 7cm},clip,angle=0]{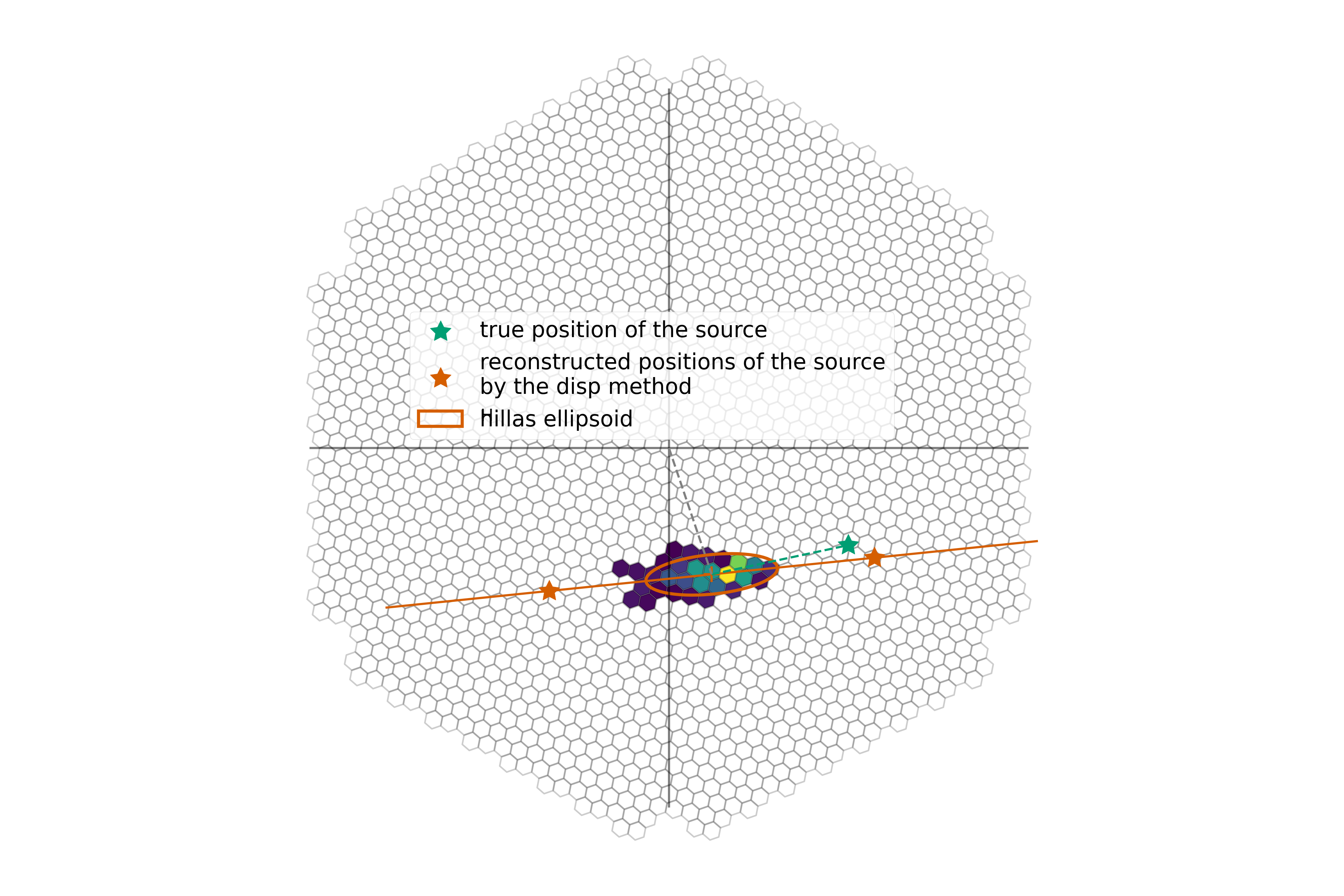}
    \caption{
    Reconstruction of the source position in the camera frame. In green the true gamma-ray direction (source position), in orange, the two possible reconstructed source positions determined by \textit{disp\_norm} along the major image axis. The final source position will be determined by \textit{disp\_sign}.}
    \label{fig:source_disp}
\end{figure}

To reconstruct the incoming direction of shower primaries, we use a version of the \textit{disp} method presented in \cite{LESSARD20011}. We assume that the point which corresponds to the event direction is located along the main image axis (as expected, within statistical fluctuations, for gamma-ray initiated showers). Two random forests models are trained using gamma MC as input: a regressor for \textit{disp\_norm} and a classifier for \textit{disp\_sign}, where \textit{disp\_norm} 
is the distance between the image centroid (x, y) and the point along the axis which is closest to the true gamma-ray direction, and \textit{disp\_sign} represents on which side of the centroid, along the major axis, the true direction lies  (see Figs.~\ref{fig:image:sketch} and \ref{fig:source_disp}). The image parameters used as input for the models are \textit{log\_intensity}, \textit{width}, \textit{length}, \textit{wl}, \textit{skewness}, \textit{kurtosis}, \textit{time\_gradient}, \textit{leakage\_intensity\_width\_2}, \textit{az\_tel} and \textit{alt\_tel}.

\subsubsection{Energy Reconstruction}
For the energy reconstruction, we use a random forest regressor trained to reconstruct the log of the true energy in TeV with the following parameters: 
\textit{log\_intensity}, \textit{width}, \textit{length}, \textit{x}, \textit{y}, \textit{wl}, \textit{skewness}, \textit{kurtosis}, \textit{time\_gradient}, \textit{leakage\_intensity\_width\_2}, \textit{az\_tel} and \textit{alt\_tel}.

\subsubsection{Particle Classification}
For the particle classification, a random forest classifier is trained with the same parameters as the disp ones, augmented with the output of the first steps: \textit{log\_reco\_energy}, \textit{reco\_disp\_norm}, \textit{reco\_disp\_sign}.\newline

\begin{figure}[ht]
    \centering
    \includegraphics[width=0.95\linewidth]{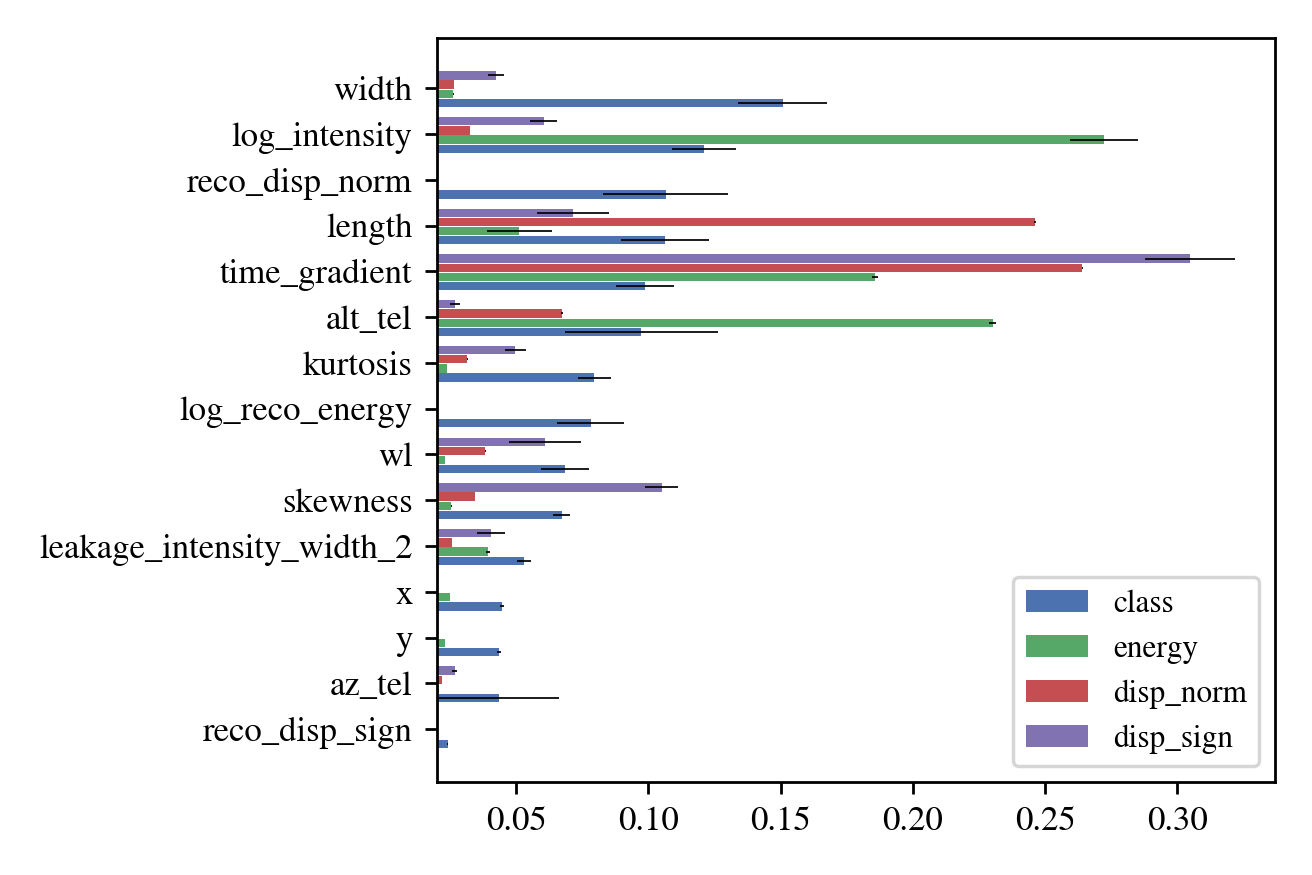}
    \includegraphics[width=0.95\linewidth]{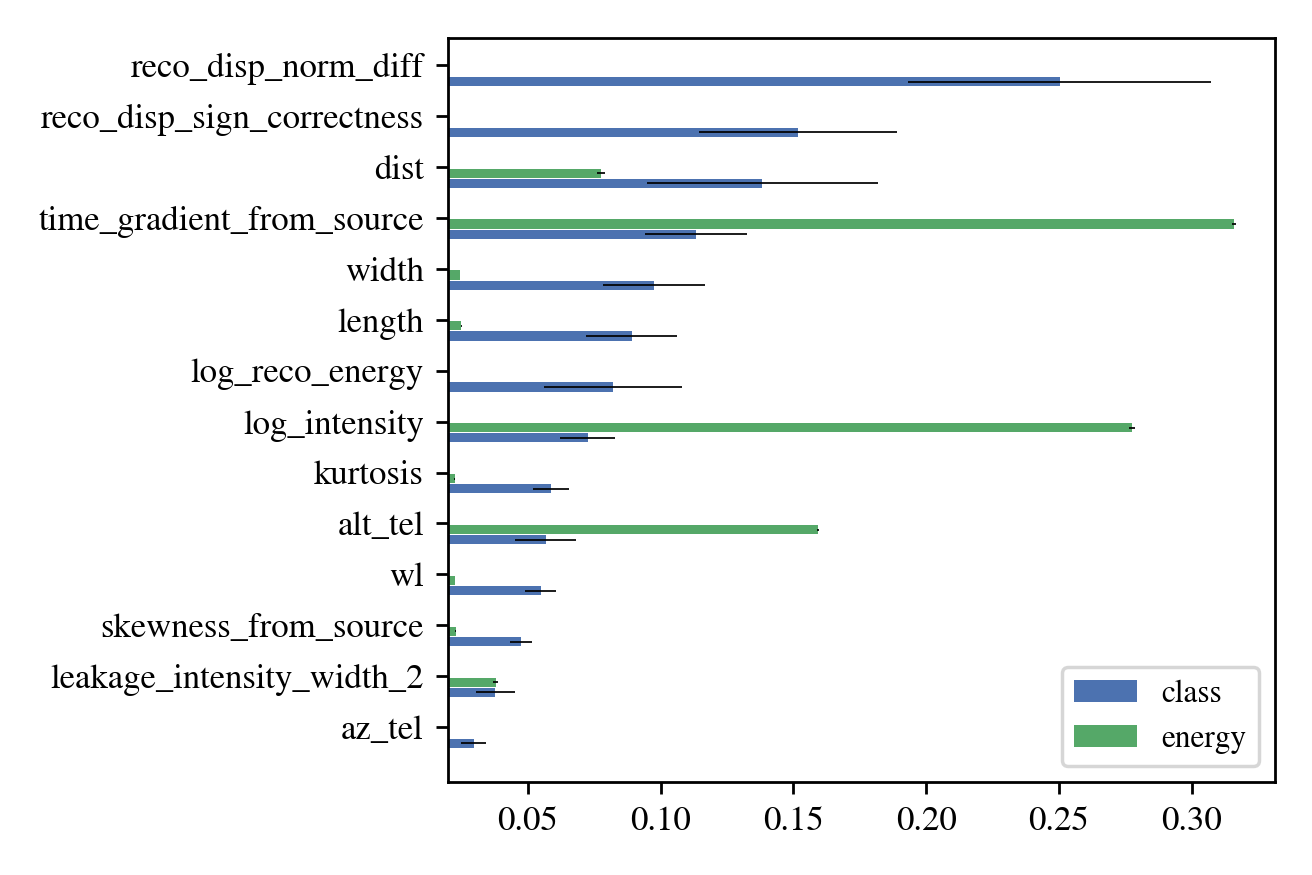}
    \caption{Relative features importance for the different random forests. Top panel is for the source-independent analysis while bottom panel is for the source-dependent analysis. Note that there are correlations among the parameters, so the true importance of a given parameter may be slightly different to the one shown here. It can also depend on the energy range: the displayed values are for the whole sample of events surviving image cleaning.}
    \label{fig:rf_imp}
\end{figure}

The relative importance of the training parameters for each model is reported in Fig.~\ref{fig:rf_imp}. It is computed using scikit-learn implementation as the mean (and standard deviation for the  error bars) of the total reduction of the Gini impurity \citep{gini1912variabilita, breiman1984classification} brought by each feature over all the trees in the random forest.
It is noteworthy that different models preferentially base their decision based on different parameters. We can also recognize the importance of the timing information for a single telescope such as \LST{}, in particular to determine the event direction. As a proxy for the impact distance, \textit{time\_gradient} is also of major importance for the energy reconstruction with a single IACT \citep{MAGIC_mono}. Note that Fig. \ref{fig:rf_imp} shows the relative importance of the parameters for the full MC training sample, considering events of all energies (and it is therefore dominated by events close to the energy threshold).

The models hyperparameters (e.g. the number and depth of the trees) have been chosen to achieve good physics performances while keeping the computing footprint manageable, one of the limitations being the memory usage during training.
The handling of the models training and production of the IRFs has been done on the collaboration computing cluster located on-site at La Palma thanks to the library \texttt{lstMCpipe} \citep{garcia2022lstmcpipe, lstmcpipev090} developed specifically for that purpose.

\subsection{Instrument Response Functions}

After training, the random forest models are applied to the MC in each of the test sample pointing nodes (see Fig.~\ref{fig:pointing_nodes}) to compute the IRFs in those directions. The IRFs are used to assess the performances of the \LST{} as a function of the energy and pointing direction, and later used for data analysis.

\subsubsection{Effective Collection Area}
The effective area is defined as the ratio of reconstructed gamma rays (after event selection cuts) over the number of simulated ones, multiplied by the area (orthogonal to the incident direction) over which events have been simulated. It is computed as a function of the true energy.

\subsubsection{Energy and Angular resolution}
With $\theta$ the angular distance between the true gamma-ray direction and the reconstructed one, the angular resolution $\theta_{68}$ is typically defined as the angle within which 68\% of the reconstructed $\theta$ values are contained. It is computed as a function of the true energy. The gamma-ray point-spread function for a single IACT has a central component made of events with properly determined head-tail image orientation (i.e. correctly reconstructed {\it disp\_sign}, see Figs. \ref{fig:image:sketch} and \ref{fig:source_disp}), and, especially at low energies, a separate tail made up by events with wrong orientation. Near threshold the fraction of correctly oriented images is less than 60\%, but it increases fast with energy (see section \ref{sec:gamma_E_distribution}). In order to characterize the central part of the PSF (which is the relevant one to show e.g. the capability of the instrument to resolve two nearby point-like sources), we consider only the population of all well-oriented MC gamma rays in the computation of $\theta_{68}$. Note that in the analysis of the observations of extended sources, or of fields with several sources, the complex shape of the PSF at low E should in principle be taken into account. In practice, however, the LST1 performance in those cases is more strongly limited by the modest background rate suppression characteristic of a monoscopic instrument.

The relative energy resolution is defined as the value of the quantity $|E_R - E_T|/E_T = |\Delta E|/E_T$ within which 68\% of the reconstructed gamma-ray events are contained, with $E_T$ the true energy and $E_R$ the reconstructed energy. The energy bias is computed as the median of $\Delta E/E_T$. Both resolution and bias are computed as a function of the true energy.

\begin{figure}[ht]
    \centering
    \includegraphics[width=1.1\linewidth]{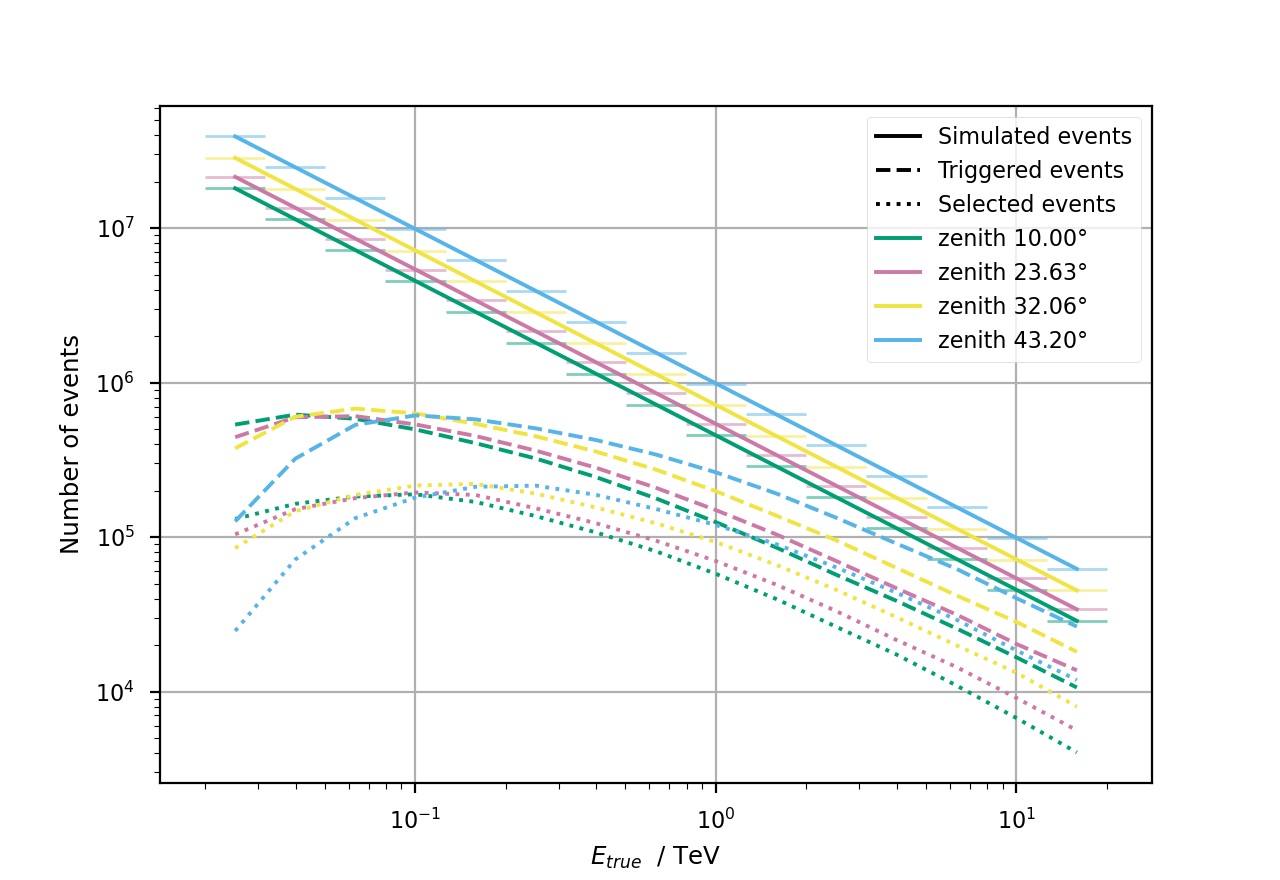}
    \caption{Number of simulated, triggered and selected events to compute the effective area and other IRFs presented in the right panel of Fig~\ref{fig:irfs_src_dep} (gamma efficiency = 0.7)}
    \label{fig:irfs_n_events}
\end{figure}

The IRFs are computed using the \texttt{pyirf} package \citep{pyirf} after some necessary event selection cuts. A global event selection $intensity  > 50$  p.e. is applied (see section \ref{sec:data_sample}), followed by an energy-dependent cut in {\it gammaness} (a minimum required value), calculated to keep a given fraction of the MC gamma rays in each bin of reconstructed energy $E_R$. And finally, for the effective area and energy resolution and bias, an energy-dependent $\theta$ cut which keeps in each bin of $E_R$ 70\% of the MC gamma rays with best direction reconstruction (among the well-oriented ones, see above). The number of simulated, triggered and selected events to produce the effective area and the other IRFs are presented in Fig~\ref{fig:irfs_n_events} for each zenith angle. It can be seen that the statistics per bin after all cuts, up to 20 TeV, are larger than several thousand events (above $10^4$ for most of the energy range), hence we expect the statistical uncertainties in the IRF computation to be typically below the percent level. Note that even though we re-use the simulated showers by detecting each one from 10 different telescope locations, after all cuts the average number of times a shower appears in the final sample is $\simeq$ 1.0, 1.4 and 2.0, below 100 GeV, at 1 TeV, and at 20 TeV respectively.

\begin{figure*}
    \centering
    \includegraphics[width=0.49\linewidth]{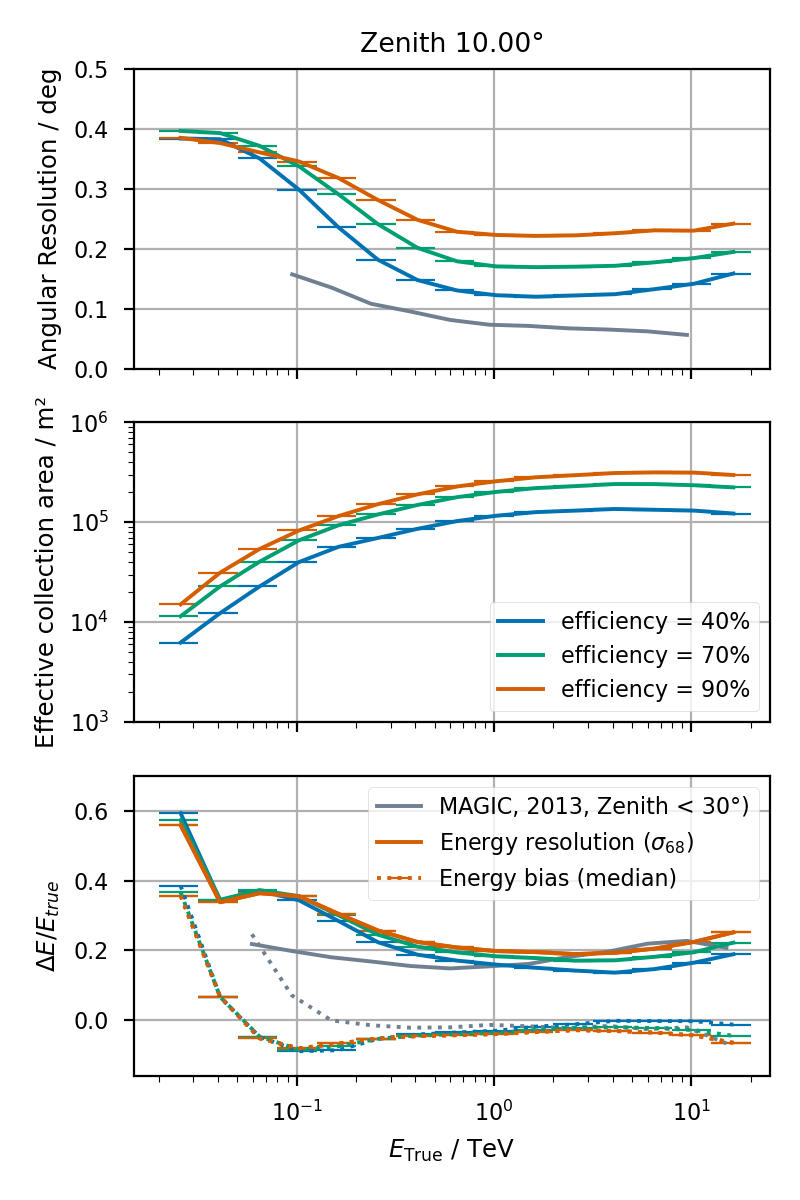}
    \includegraphics[width=0.49\linewidth]{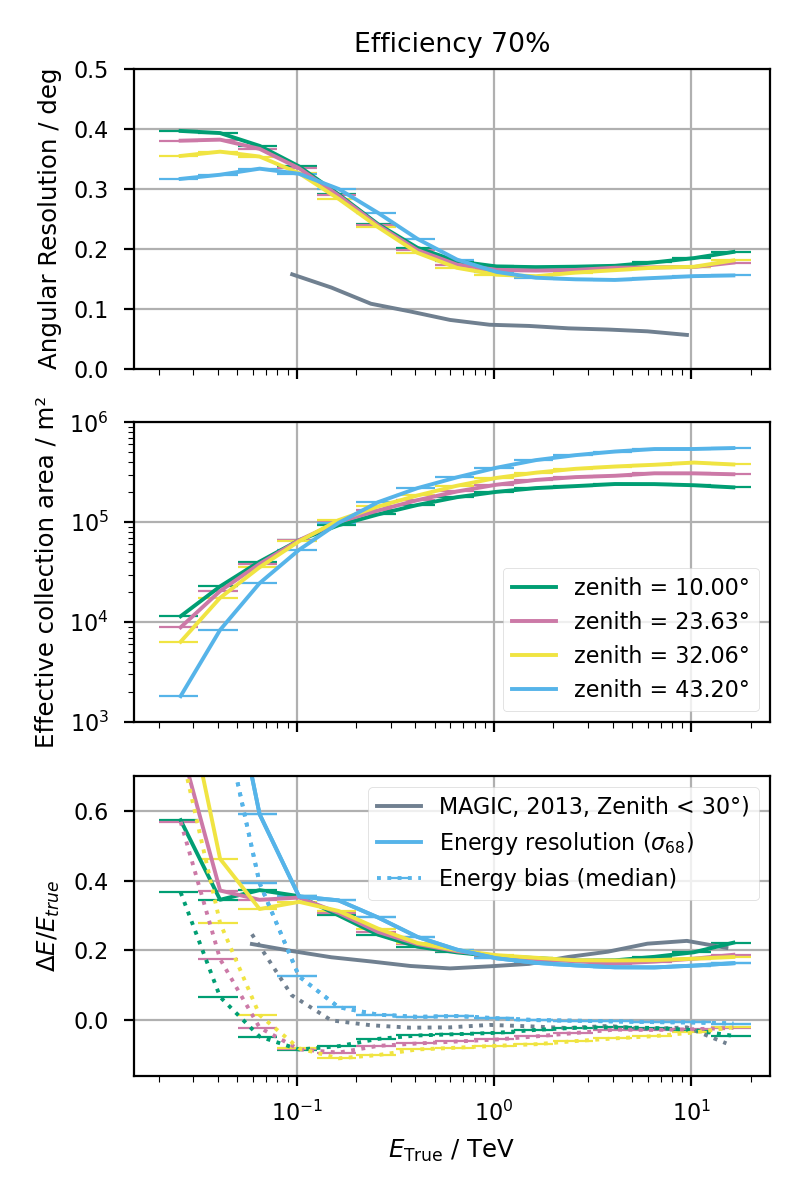}
    \vspace{-0.3cm}
    \caption{IRFs as a function of the true energy. Left panels: fixed Zenith = $10\degree$ and several gamma-ray efficiencies, right panels: fixed gamma-ray efficiency = 70\% for several zenith angles. 
    Top panels: the angular resolution, mid panels: the effective area, bottom panels: the energy resolution and bias.
    Angular and energy resolution are best at intermediate energies, worsening towards high energies due to the truncation of the large-impact shower images, and towards low energies due to the less precise reconstruction of small and dim showers. The performances of MAGIC extracted from \cite{ALEKSIC201676} are shown for comparison.
    }
    \label{fig:irfs_srcindep}
\end{figure*}

The resulting IRFs are presented in Fig.~\ref{fig:irfs_srcindep}. The left panels provide them for a zenith angle of $10 \degree$ and for several efficiencies of the {\it gammaness} cut. As one can expect, the reconstruction performance generally improves (better resolutions and lower bias) with lower efficiencies (due to stricter event selection), at the expense of collection area. An exception to that rule can be seen in the energy resolution near threshold, possibly because in that range a tighter gamma-ray selection cut may also entail the selection within an $E_T$ bin of up-fluctuations in terms of e.g. Cherenkov light yield, leading to a larger energy reconstruction bias.

The best performances at low zenith ($10 \degree$) are achieved in the TeV range, with an angular resolution between $0.1\degree$ and $0.2 \degree$ depending on the gamma-ray selection efficiency, a relative energy resolution down to 15\%, and a bias of 5\%. Below 100~GeV, performances decrease rapidly but stay relatively constrained at low zenith. The right panels of Fig.~\ref{fig:irfs_srcindep} present IRFs for a fixed gamma-ray selection efficiency of 70\% and  for zenith angles between $10\degree$ and $43.2\degree$, close to the zenith angles of the observations presented in the next section. We currently have no explanation for the "bump" in the energy resolution curves (located around 70 GeV for low zenith), but it seems a genuine feature: it has a smooth structure when finer E-binning is used, and it appears, slightly shifted in energy, at different zenith angles.

\begin{figure*}
    \centering
    \includegraphics[width=0.49\textwidth]{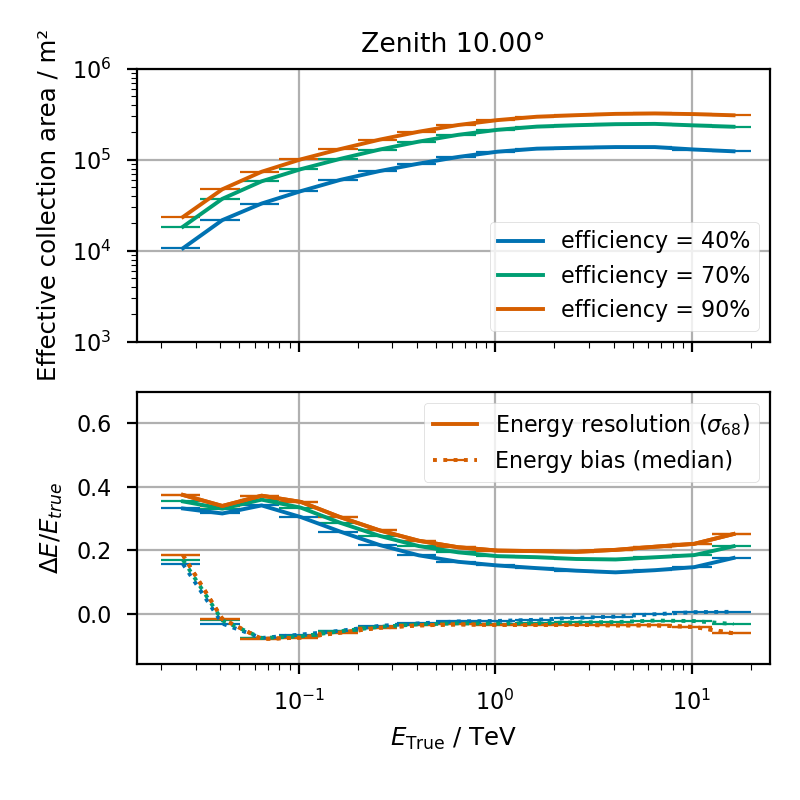}
    \includegraphics[width=0.49\textwidth]{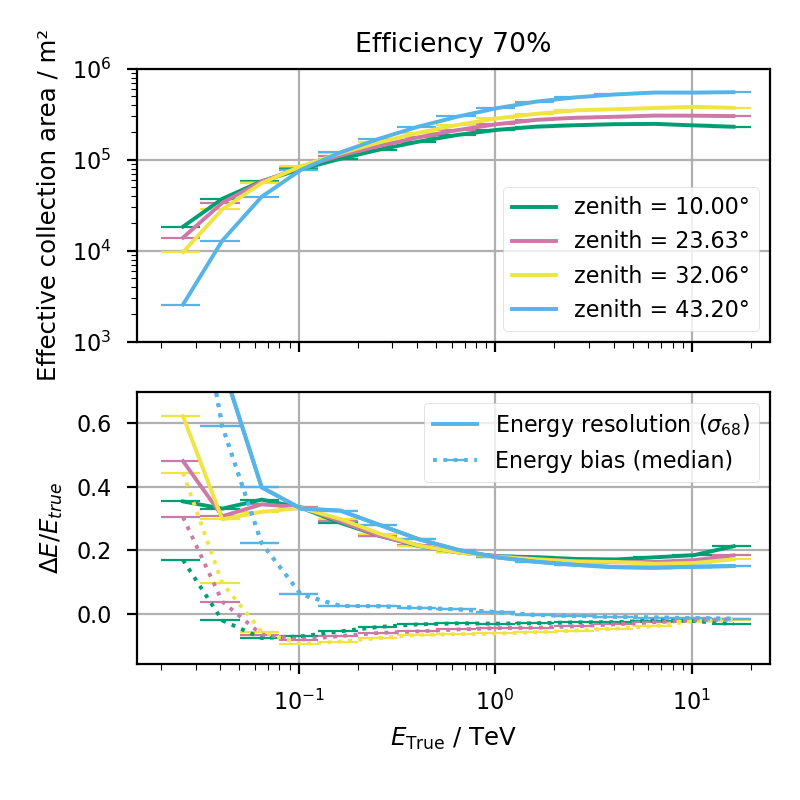}
    \vspace{-0.3cm}
    \caption{Evolution of the source-dependent analysis IRFs as a function of the true energy. Left panel: fixed Zenith = $10\degree$ and several gamma-ray efficiencies. Right panel: fixed gamma-ray efficiency = 70\% and several zenith angles. Top panels: the effective area. Bottom panels: the energy resolution and bias. The energy resolution shows the same general trend as for the source-independent analysis, but with a significantly better value in the first bin.}
    \label{fig:irfs_src_dep}
\end{figure*}

\subsection{Source-dependent Analysis} 
The event reconstruction performance using a single telescope image is expected to be worse than the stereoscopic reconstruction, especially in the low-energy range. To improve the monoscopic analysis performance, an a priori assumption of the gamma-ray source position may be advantageous. In the case of a single point-like gamma-ray source in the telescope field of view, all gamma rays are expected to arrive from the same direction. A powerful parameter used in the source-dependent analysis is \textit{dist} (see Fig.~\ref{fig:image:sketch}), which is the distance between the known source position and the centroid of the shower images. Since \textit{dist} correlates with the shower impact parameter inside the light pool~\citep{MAGIC_mono}, it improves the energy reconstruction performance. 

For source-dependent analysis, the signs of \textit{skewness} and \textit{time\_gradient} are redefined based on the known source position, and the parameters are renamed \textit{skewness\_from\_source}, \textit{time\_gradient\_from\_source}. Those source-dependent parameters including \textit{dist} are used as input parameters of the random forest training. For proton MC, we use a single fixed point 0.4$^\circ$ away from the camera center to compute the source-dependent parameters. Thus, the image centroid coordinates (\textit{x}, \textit{y}) are removed as training input parameters for source-dependent analysis to avoid bias.

For the particle classification, two extra parameters ({\it reco\_disp\_norm\_diff}  and {\it reco\_disp\_sign\_correctness}) are introduced based on the comparison between the direction reconstruction and the known source position. The former is the absolute value of the difference between {\it reco\_disp\_norm} and {\it dist}, and the latter is the estimated probability (see the Appendix) that the known source position is the correct one for the given image. Both are measures of how consistent the result of the \textit{disp} method is with the known position of the source. The input parameters used for the source-dependent analysis are also shown in the bottom panel of Fig.~\ref{fig:rf_imp}. It can be seen that the four most relevant parameters for image classification use the knowledge of the source position.

Besides including the additional parameters, in the random forest training we only use gamma-ray simulation events with incident direction within 1.0$^{\circ}$ of the telescope pointing (a compromise between keeping good training statistics while excluding events with significantly larger off-axis angles than the Crab has in the real observations).

To compute the excess counts, the \textit{alpha} angle (the angle between shower axis and the line between the known source position and the image centroid, see Fig. \ref{fig:image:sketch}) is used instead of $\theta$ for the source-dependent analysis. 
The IRFs for source-dependent analysis are presented in Fig. \ref{fig:irfs_src_dep}, again for several gamma-ray efficiencies and zenith angles. In this case, the angular resolution cannot be computed because the direction of gamma-ray events is assumed to be known. 
The applied selection cuts for the IRFs are then: $intensity$ > 50 p.e., the gamma-ray efficiency and an additional $\alpha\textrm{-cut efficiency}=70\%$ (similar to the $\theta$-efficiency) for the effective area, energy resolution and energy bias. The obtained IRFs are very similar to those of the source-independent analysis, the main difference being a visible improvement of the energy reconstruction in the first two energy bins (i.e. true energy below 50 GeV). There is also an increased effective area at low energies (by around 40\% under 50 GeV) - due to events that in the source-independent analysis are reconstructed far away from the source because of wrong head-tail assignment.

\subsection{Energy Distribution of Gamma Rays in Different Analysis Stages \label{sec:gamma_E_distribution}}
\begin{figure*}
\begin{center}
\includegraphics[width=\textwidth]{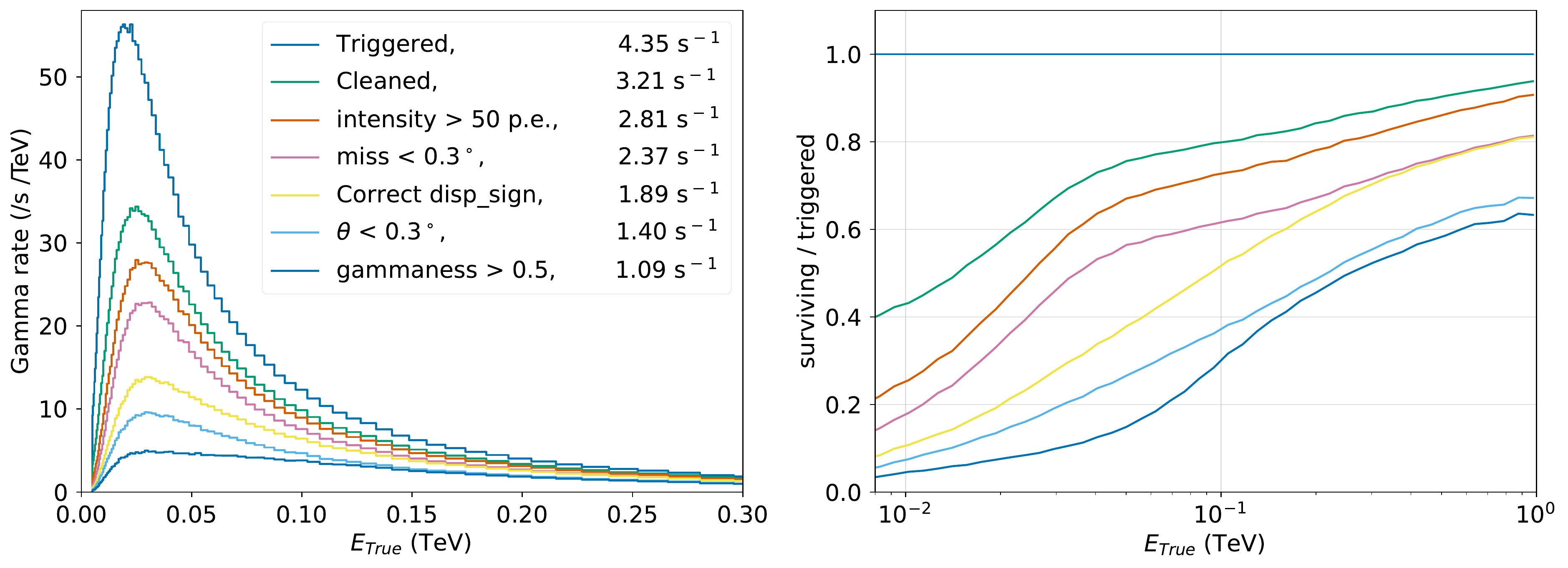}
\vspace{-0.3cm}
\caption{Left: Distribution of the true energy of MC gamma rays observed at 10$\degree$ zenith angle, at different stages of the standard source-independent analysis. Energy-dependent weights are applied to the events to reproduce a Crab-like spectrum (from \cite{CrabMAGIC}), resulting in the quoted event rates. The peak energy for triggered events (often used as a definition of energy threshold) is at $\simeq 20$ GeV. The peak shifts to $\simeq 25$ GeV after cleaning, and $\simeq 30$ GeV when we require a minimum image \textit{intensity} and exclude events with bad reconstruction of the image axis (miss is the distance between the true gamma-ray direction on the camera and the reconstructed axis, see Fig. \ref{fig:image:sketch}). Right: fraction of triggered events which survive the different stages of the analysis, plotted vs. energy. Below 100 GeV, due to the limitations of monoscopic shower reconstruction, a majority of the surviving gamma-ray events have poorly reconstructed direction and/or are hard to distinguish from the background.}
\label{fig:energy_threshold}
\end{center}
\end{figure*}

The energy threshold of an IACT is often defined as the energy at which the distribution of the true energies of the detected events peaks, for a given assumed spectrum (usually that of the Crab Nebula). Fig. \ref{fig:energy_threshold} shows the $E_\text{True}$ distributions for low-zenith (10~$\degree$) MC gamma rays at trigger level and at different stages of the analysis. The simulations are tuned to the current trigger configuration of the telescope (see section \ref{sec:data_sample}). The trigger threshold is about 20 GeV, which rises to $\simeq 30$ GeV after selecting the events which survive the cleaning stage, have an image \textit{intensity} of at least 50 p.e., and have a well-reconstructed image axis (closer than 0.3$^\circ$ to the true direction). Naturally, the loss of events in all of the analysis steps is larger near threshold than it is at higher energies. Below 100 GeV, nearly half of the events for which a reasonably {\it well-oriented} image main axis has been reconstructed get a poor direction reconstruction because of wrong head-tail assignment ({\it disp\_sign}), or poor {\it disp\_norm} reconstruction. This just means that for faint, few-pixel images, determining the plane which contains the shower axis and the telescope location (except at very low impact parameter) is much easier than determining the direction of the axis within that plane. That is the main reason why stereoscopic reconstruction enormously improves the performance of IACTs: additional images of the same shower provide more such planes, from whose intersection a full 3D geometrical reconstruction of the shower axis can be achieved. It is also the reason why source-dependent analysis results in improved energy resolution for a single IACT.

Fig. \ref{fig:energy_threshold} also illustrates the fact that, for a fixed {\it gammaness} cut, the loss of gamma-ray events is always larger at lower energies, i.e. the background discrimination power in monoscopic analysis decreases fast as the images become fainter. As we will see, the lack of stereoscopic reconstruction makes that despite its large mirror area and lower energy threshold, \LST{} operating in monoscopic mode cannot outperform (in the overlapping energy range) the existing arrays of IACTs with significantly smaller mirror dishes.  

\section{The data sample} \label{sec:data_sample}
\begin{figure*}
\begin{center}
\includegraphics[width=\textwidth]{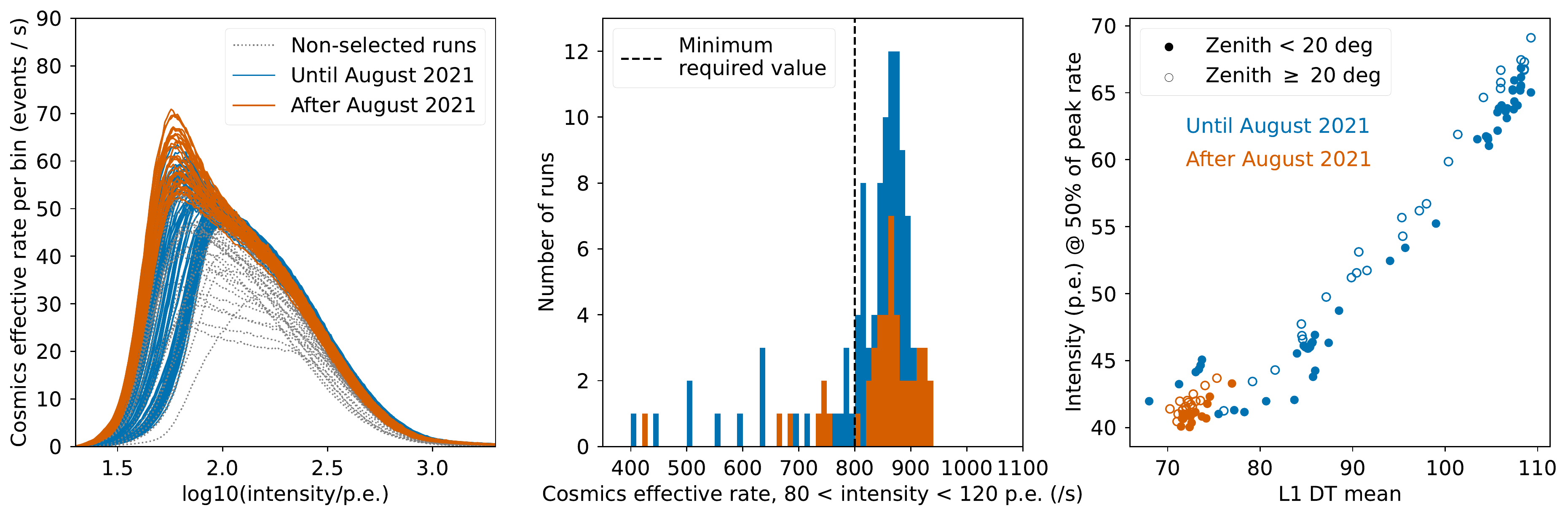}
\vspace{-0.3cm}
\caption{Left: run-wise image \textit{intensity} distributions (for shower events). Each curve corresponds to a data run. The large differences seen below 80 p.e. among the selected runs are related to the different trigger settings of the telescope, which did not become stable until August 2021. The differences in the peak rates among the post-August 2021 data are mostly connected to the different zenith distances, which span the range $6\degree$ to $35 \degree$. 
Center: rate of cosmics in the \textit{intensity} range between 80 and 120 p.e. The runs with a value below 800 s$^{-1}$ are discarded.
Right: correlation between the position of the rising edge of the \textit{intensity} distributions and the camera-averaged trigger threshold setting. The deviation from the linear behavior at low thresholds is related to the image cleaning, which removes the faintest among the triggered events.}

\label{fig:intensity_spectra}
\end{center}
\end{figure*}

The data used in this study were recorded between November 2020 and March 2022. We have focused on the analysis of low-zenith Crab observations performed in wobble mode \citep{FOMIN1994137}, which facilitates the task of estimating the rate of background events recorded together with the signal. The telescope was pointed $0.4 \degree$ away from the center of the Crab Nebula, alternating between two different sky directions on opposite sides of the source. Each observation run with a given sky pointing lasted typically for 20 minutes. A total of 57.2 hours of observation with the telescope pointing within $35 \degree$ of the zenith were collected. Of those, 48.0 hours correspond to dark night observations, i.e. with the Moon below the horizon. This is our starting sample. The zenith and dark-night constraints are aimed at achieving a low gamma-ray energy threshold of around 20 GeV. The next step was to identify the data taken under good atmospheric conditions. The total shower trigger rates are affected by weather conditions (which modify the atmospheric transmission), but also by variations in the telescope trigger settings. During this period, the telescope was in its commissioning phase, and different trigger settings were tested (including different algorithms for the dynamic modification of the settings during observations, to react to the presence of stars in the field of view). This means that the threshold of \LST{} was not stable (see right panel of Fig. \ref{fig:intensity_spectra}), and hence the {\it total} shower trigger rates are not a good proxy of the quality of the atmospheric conditions during this period. Instead of total rates, we used two quantities which are not affected by the trigger settings, because they are determined by showers well above the threshold: the camera-averaged rate of pixel pulses with charge above 30 p.e. ({\it pix\_rate} $_{q>30}$), and the rate of shower images with \textit{intensity} between 80 and 120 p.e. ($R_{80-120}$) (see the central panel of Fig. \ref{fig:intensity_spectra}). We calculated the run-averaged values for those quantities, and removed from our sample runs with {\it pix\_rate} $_{q>30}$ < 4.5 s$^{-1}$ or $R_{80-120}$ < 800 s$^{-1}$. The cuts are just intended to remove outliers, and the specific cut values are to a certain extent arbitrary. The total observation time of the 117 surviving runs amounts to 35.9 hours. Using the distribution of time intervals between consecutive triggered events, we estimated the dead time to be around 4.7\%, resulting in an effective observation time of 34.2 hours.

The left panel of Fig. \ref{fig:intensity_spectra} shows the run-averaged distributions of shower image intensities. The grey dotted lines correspond to runs that were rejected by the cut in $R_{80-120}$. The rest of the distributions are those of the selected runs. Above $\simeq$80 p.e. all the spectra agree pretty well, with the maximum and the minimum rates differing by less than 20\% (with part of this spread being actually due to the different zenith angles). Below 80 p.e., in contrast, large differences appear, caused by the variations in the trigger settings discussed above. We can characterize the threshold of each of these distributions by the value of $\log_{10}(intensity/\text{p.e.})$ at which 50\% of the peak event rate is reached.  The right panel of Fig. \ref{fig:intensity_spectra} displays that parameter (converted to p.e.) as a function of the camera-averaged trigger threshold setting during the run: the plot clearly shows that the differences among the \textit{intensity} distributions are indeed mostly due to the different trigger settings. The current configuration of the trigger settings was established in August 2021. The data taken since then (see orange curves in Fig. \ref{fig:intensity_spectra}, left) are much more stable and have a lower threshold.

The trigger threshold in the MC simulations has been set to the lowest value found in the data, i.e. it is tuned to the post-August 2021 situation. This means that in order to achieve a good match between the full data sample and the simulations, we have to apply a \enquote{software trigger} to equalize data and MC. In this analysis we apply a simple cut in image \textit{intensity}. A cut $intensity > 80$ p.e. brings all of the data to a common \enquote{analysis threshold} and ensures a good match of MC and data. For data taken after August 2021, a cut $intensity > 50$ p.e. is enough to achieve the same goal.

\section{Validation of the MC simulation through comparisons with data}
\label{sec:data_mc_comparison}

We have used the observations of the Crab Nebula to validate the Monte Carlo simulation of the detection of gamma rays with \LST{}. We do this through the comparison of the distributions of image parameters for the simulated gamma-ray images, and those obtained for the gamma-ray excess recorded from the direction of the Nebula. For a meaningful comparison, the distribution of the energies of the simulated and observed gamma rays must be as close as possible. The MC histograms have been filled with event-wise energy-dependent weights calculated to reproduce the log-parabola parametrization of Crab Nebula spectrum reported in \cite{CrabMAGIC}. For better comparison of the distribution shapes, the overall normalization of the MC histograms is a free parameter, tuned to achieve the best match between the distributions (via a $\chi^2$-test). The small extension of the Crab Nebula \citep{2020NatAs...4..167H} is also simulated using a Gaussian smearing ($\sigma_{2D} = 52.2''$) of the reconstructed directions in the MC sample. The comparisons are presented in four ranges of image \textit{intensity}, starting at 80 p.e.: 80-200, 200-800, 800-3200 and >3200 p.e. The distributions of true (MC) energy peak respectively at $\simeq$ 30, 80, 200 and 700 GeV (see right panel of Fig. \ref{fig:ROC_and_E}).

\begin{figure*}
\begin{center}
\includegraphics[width=\textwidth]{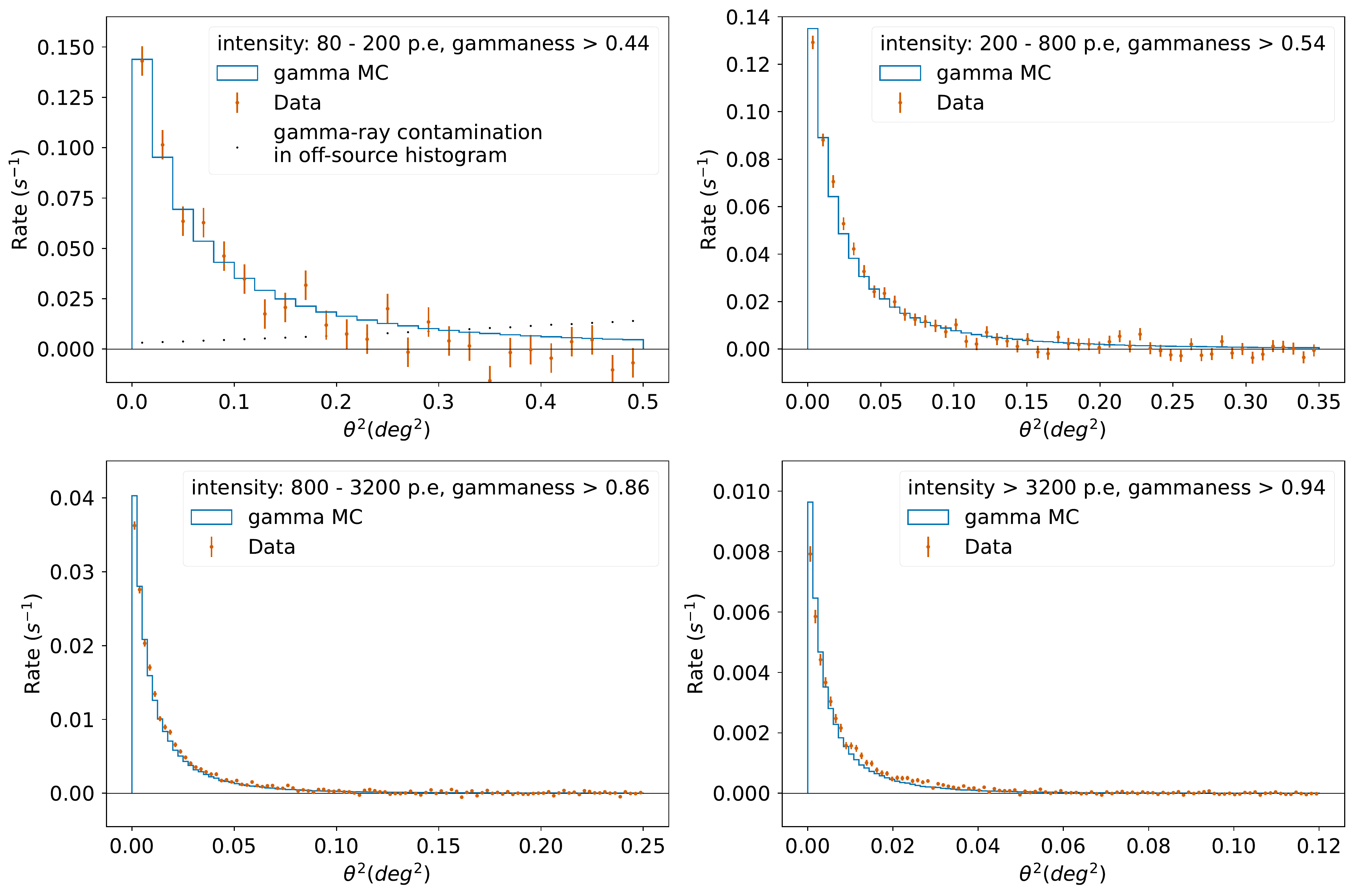}
\vspace{-0.3cm}
\caption{Comparison of $\theta^2$ distributions, gamma MC simulations vs. Crab Nebula excess events. The observed discrepancies may be partly due to arcminute-scale mispointing of the telescope.}
\label{fig:MC_vs_data_theta2}
\end{center}
\end{figure*}

We start by comparing the angular distribution of the events around the center of the Nebula. Fig.~\ref{fig:MC_vs_data_theta2} shows the distribution of the $\theta^2$ parameter, the squared angular distance between the reconstructed event directions and the source. The distribution for the gamma rays in the real data is obtained by subtracting the distribution calculated with respect to a control \enquote{off-source} direction from the \enquote{on-source} distribution in which $\theta$ is computed relative to the Crab. The off-source direction is 0.8$^\circ$ away from the source direction, with the center of the field of view located exactly between them. The acceptance for background events (mostly proton-initiated showers) is the same around both directions at better than the $1\%$ level, and therefore the subtraction of the two distributions yields the distribution of $\theta^2$ for the gamma-ray excess. The distributions are obtained with the events which survive a gamma-ray selection (via a {\it gammaness} cut) which keeps in each \textit{intensity} range $\simeq 80\%$ of the gamma-ray rate. The cut reduces the background rate, and hence the fluctuations in the subtracted histograms. 

In the lowest \textit{intensity} range  (80 - 200 p.e.) the tail of the gamma-ray distribution extends beyond the center of the field of view, and a small correction is necessary to account for the expected gamma-ray contamination in the bins of the off-source histogram. The correction is shown by the black dots on the top left panel of Fig. \ref{fig:MC_vs_data_theta2}. As expected, the quality of the direction reconstruction improves significantly as we select brighter images. The test also shows that the MC distributions are generally narrower than those of real data. The difference becomes more noticeable as \textit{intensity} increases and angular resolution improves. This hints at the possibility that the difference between data and MC is mostly due to arcminute-scale variable mispointing (note that no offline pointing corrections have been applied in this analysis). We tested this hypothesis by introducing in the MC simulation a Gaussian smearing of the reconstructed directions. Excellent agreement of the $\theta^2$ distributions on all four ranges of intensity was achieved for a smearing with standard deviation $\sigma$ = 1.5 arcminutes in each axis. This possible mispointing is much smaller than the angular resolution achieved even for the brightest images, and hence poses no major challenge for the higher-level analysis. We expect most of this discrepancy to disappear once we implement offline pointing corrections based on the observation of stars in the field of view.

\begin{figure*}
\begin{center}
\includegraphics[width=\textwidth]{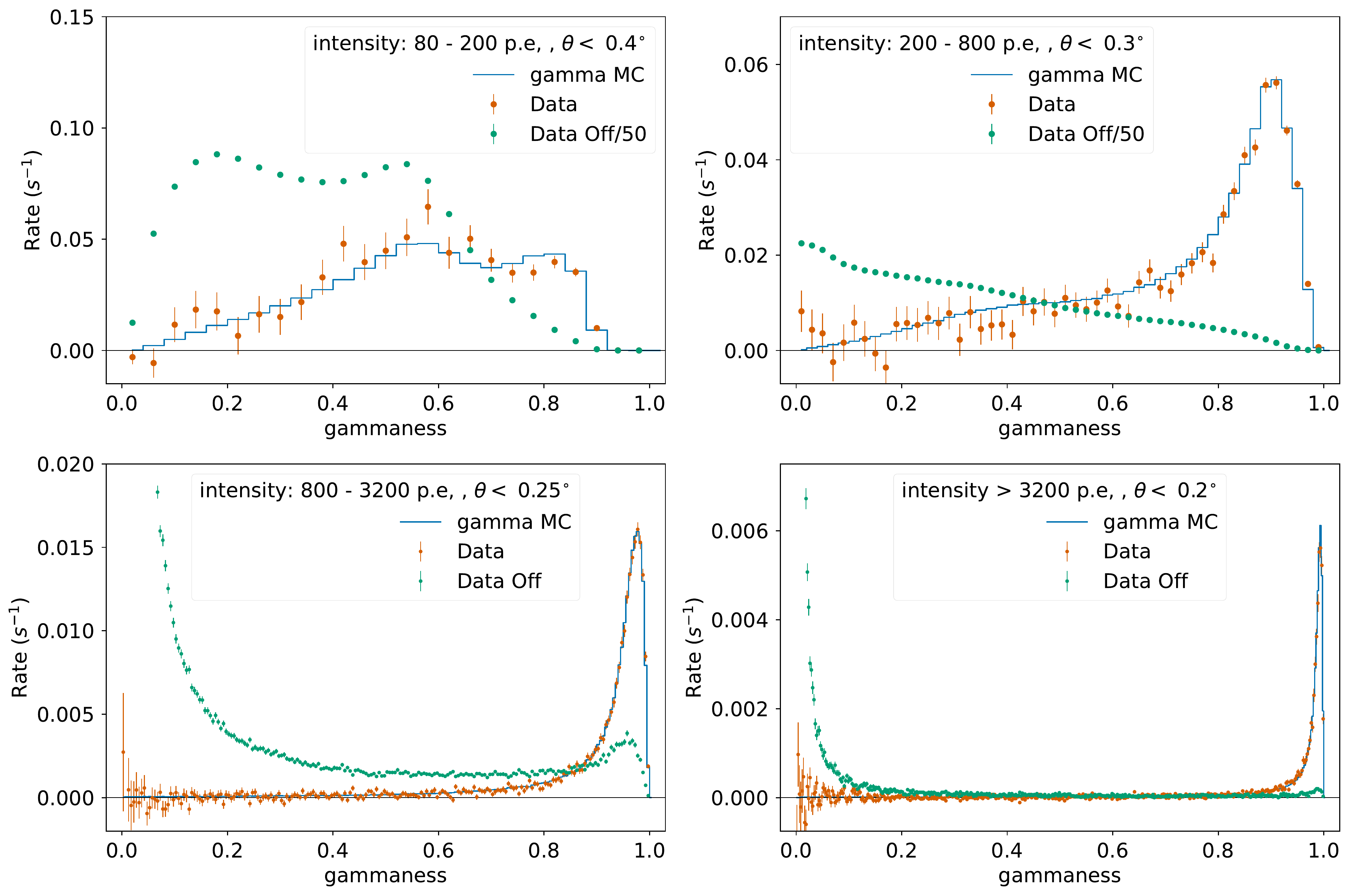}
\vspace{-0.3cm}
\caption{Comparison of \textit{gammaness} distributions, gamma MC simulations vs. Crab Nebula excess events. The improvement of the background discrimination power with the \textit{intensity} of the images is clearly seen. Note that in the top panels the background rates are scaled by a factor 1/50 for better visibility.}
\label{fig:MC_vs_data_gammaness}
\end{center}
\end{figure*}

\begin{figure*}
   \centering
    \begin{tabular}{cc}
        \includegraphics[width=0.49\textwidth]{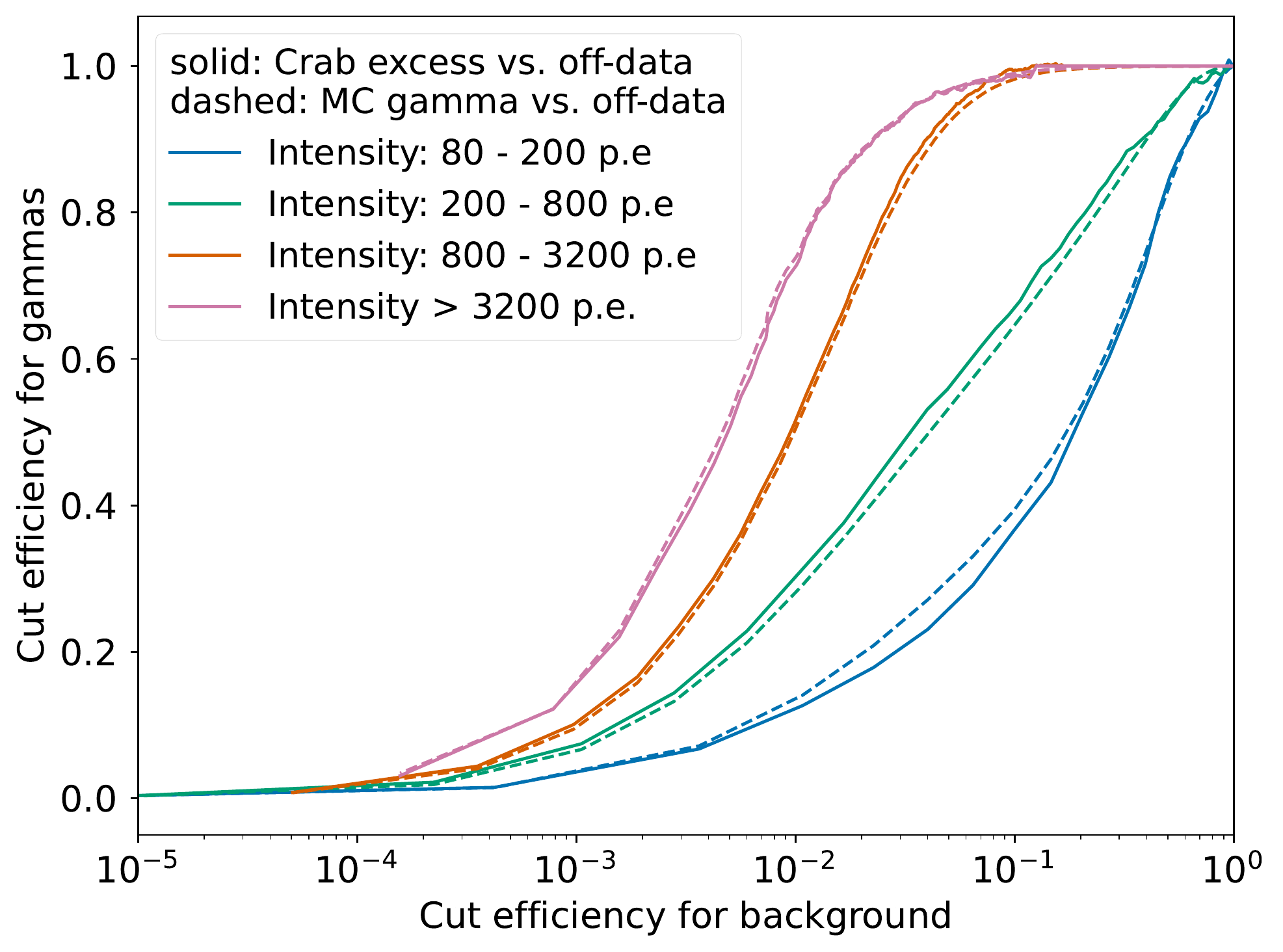} &
        \includegraphics[width=0.49\textwidth]{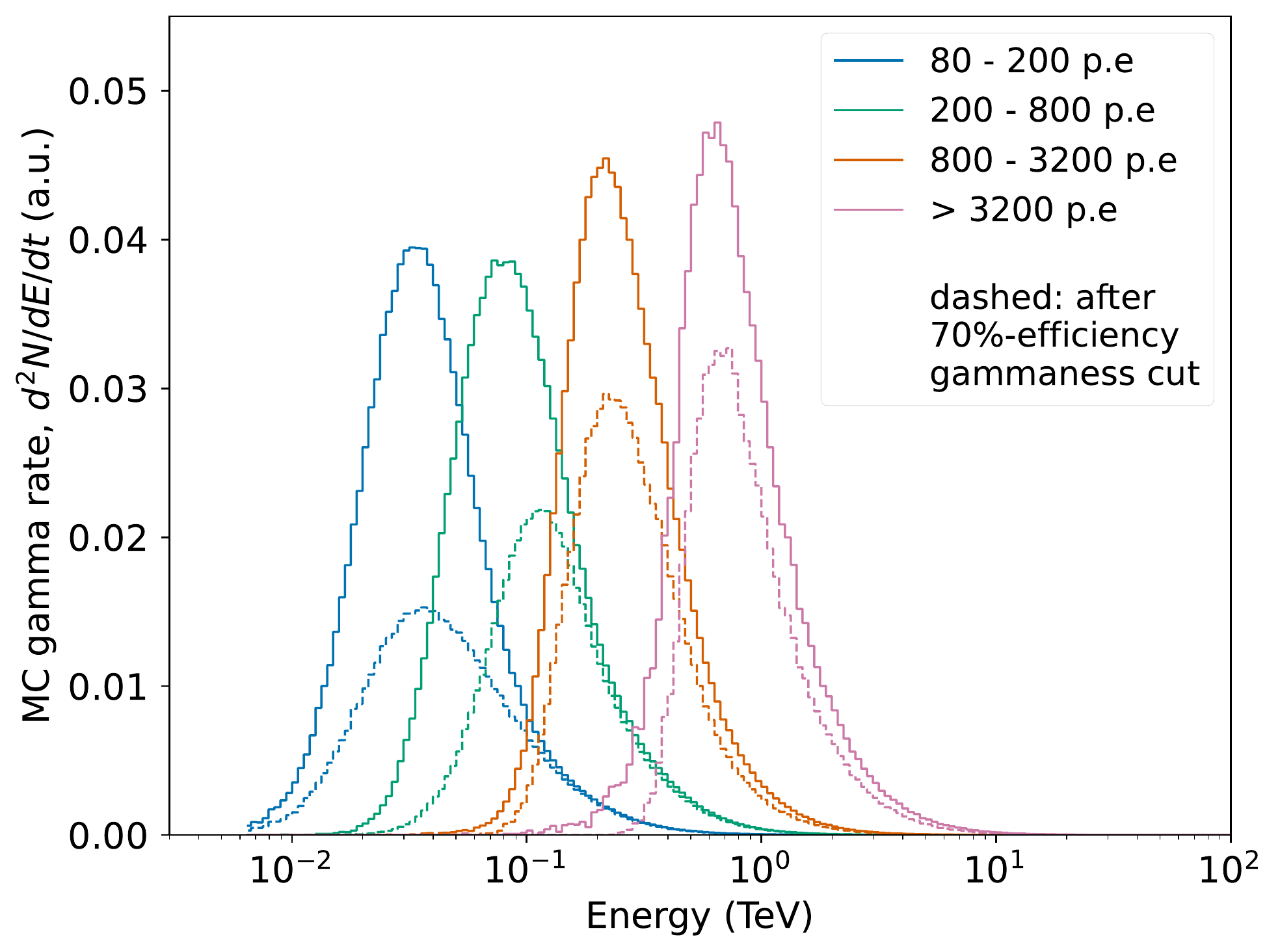}
    \end{tabular}
    \caption{Left: ROC curves of the signal selection cuts. Right: Energy distributions of gamma rays in different ranges of image \textit{intensity}. The $\theta$ cuts are the same as in Fig. \ref{fig:MC_vs_data_gammaness}. Note that the 70\% cut efficiency is computed relative to the total gamma rate in each \textit{intensity} bin (which is not proportional to the areas under the curves in this representation with logarithmic energy axis).}
    \label{fig:ROC_and_E}
\end{figure*}

For other parameter comparisons it is important to avoid the bias in the distributions that might be caused by the application of analysis cuts, like e.g. the {\it gammaness} cut described above. The distribution of any image parameter which is used as an input in the random forest which computes {\it gammaness} would be biased (towards looking more MC-gamma-like) if a cut in {\it gammaness} was applied to select the events used in the procedure. Obviously, if the parameter is {\it gammaness} itself, such a cut would simply clip the distribution. For the following parameter comparisons we therefore use {\it all} events with reconstructed direction within a given angular distance of the source to produce the distribution of the desired parameter (\enquote{on-source}). The angular cuts for the four \textit{intensity} bins ($\theta <$ $0.4\degree$, $0.3\degree$, $0.25\degree$ and $0.2 \degree$) integrate over $90\%$ of the observed excess. Due to the lack of background rejection, the on-source histograms will contain many more background events than gamma-ray events from the source. We then obtain the corresponding distribution for background-only events, using those recorded around the off-source direction, with the same angular cuts. Like in the case of $\theta^2$, the subtraction of the off-source distribution from the on-source one yields the distribution of the given parameter for the gamma-ray excess events. The result of this procedure applied to the {\it gammaness} parameter is shown in Fig. \ref{fig:MC_vs_data_gammaness}. The data and MC distributions agree well in all four \textit{intensity} bins. The distribution of {\it gammaness} for the background events is also plotted, clearly showing the increasing background discrimination power of the {\it gammaness} parameter as image \textit{intensity} gets larger. This can be better seen in the receiver operating characteristic (ROC) curves shown in the left panel of Fig. \ref{fig:ROC_and_E}. The curves are produced using real off data as background, and the observed Crab excess (solid lines), or the MC gamma rays (dashed lines), as signal. This shows once more that MC simulations reproduce well the actual behavior of \LST{}.

\begin{figure*}
\begin{center}
\includegraphics[width=\textwidth]{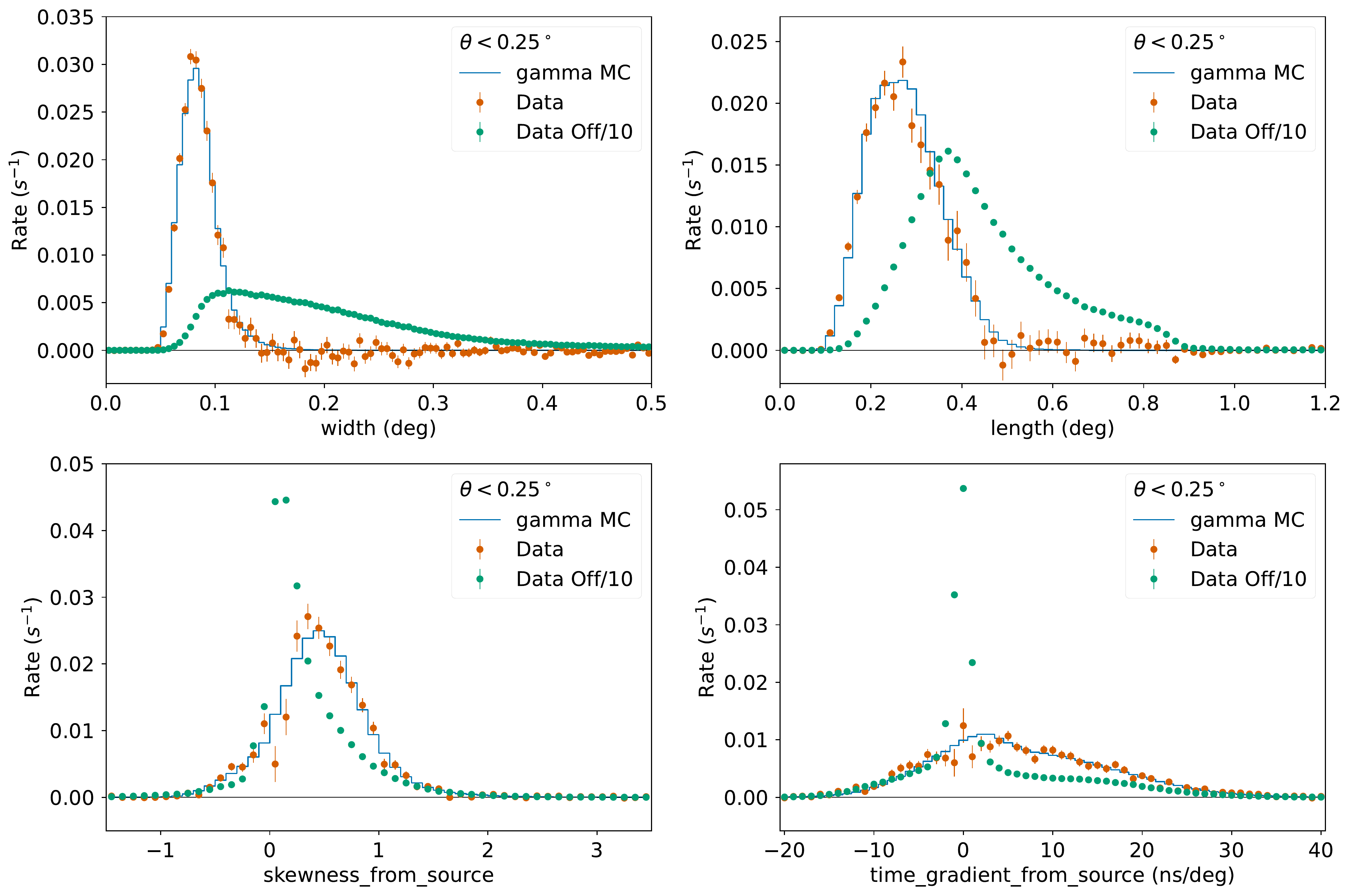}
\vspace{-0.3cm}
\caption{Distribution of several image parameters for events in the \textit{intensity} range 800 - 3200 p.e., gamma MC simulations vs. Crab Nebula excess events. The sharp peak at 0 in the time gradient distribution of the background (bottom right panel) is mostly due to events dominated by single muons. In the bottom plots, the sign of the skewness and time gradient parameters is defined relative to the true source position, to show the asymmetry that allows to determine the head-tail orientation of the shower images.}
\label{fig:Four_params}
\end{center}
\end{figure*}

The good data - MC agreement of the {\it gammaness} distributions indicates that the distributions of the individual image parameters (and the correlations among them) that are used in the computation of {\it gammaness} are also well reproduced by the simulation. We illustrate this in Fig. \ref{fig:Four_params}, which shows the distribution of four important image parameters obtained in the \textit{intensity} range 800 - 3200 p.e. (in which the distributions of individual parameters for gamma rays and background already differ significantly). 

\section{Results \label{sec:results}}

In this section we present the results obtained from the higher-level analysis of the \LST{} observations of the Crab Nebula and pulsar. These include the spectral energy distribution and light curve of the Nebula, a phaseogram showing the detection of the pulsar, and an estimate of the \LST{} flux sensitivity for point-like sources.

\subsection{Flux Sensitivity \label{sec:sensitivity}} 

To calculate the sensitivity of \LST{}, we follow the definition of detection sensitivity as the minimum flux from a point-like source that the telescope is able to detect with a 5-$\sigma$ statistical significance in 50 hours, calculated in five logarithmic energy bins per decade and using an ON/OFF ratio of 0.2. Additionally, we require at least 10 detected gamma rays per energy bin, and a signal-to-background ratio of at least 5\%. 

The calculation presented here is based on the Crab observations, and hence the result corresponds to the average \LST{} performance at low zenith angles (< 35$^\circ$). We divided the Crab data sample into two subsets of the same size (even- and odd-numbered events) which are completely equivalent in terms of telescope pointings and all other observation conditions. One of the subsets was used to optimize the gamma-ray selection cuts, by scanning a grid of $\alpha$/$\theta$ and gammaness cuts, and selecting the cut combination that results in the best sensitivity for each of the energy bins. The optimal cuts were then applied to the other (statistically independent) subset, on which the minimum flux that fulfills all conditions in the definition was computed. This procedure ensures that the sensitivity values are not biased by statistical fluctuations. In the cut scan, only cut combinations which produced at least a 3-$\sigma$ excess from the Crab direction in both subsets are considered.

\begin{figure}
    \centering
    \includegraphics[width=0.49\textwidth]{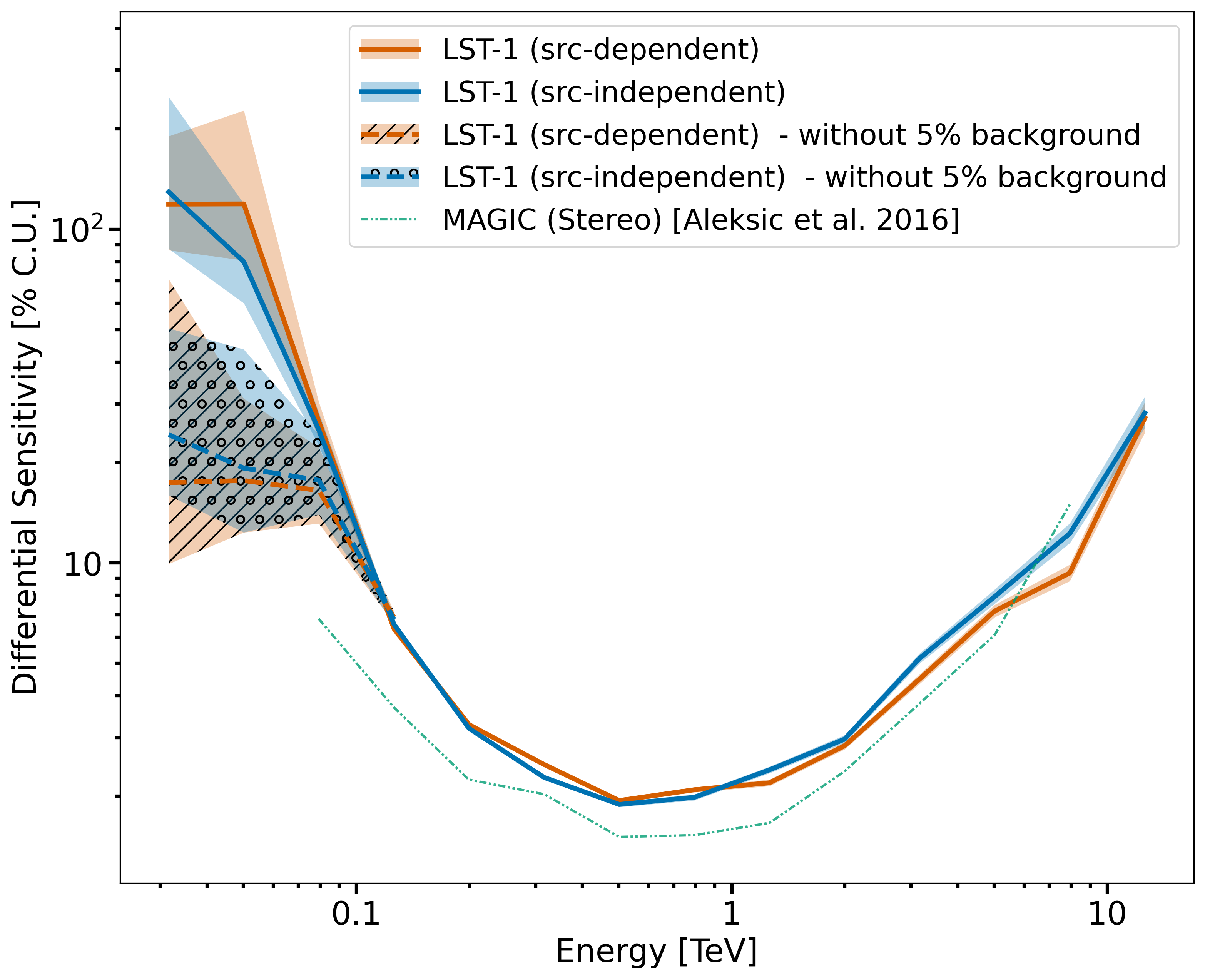}
    \caption{Differential sensitivity for source dependent and source independent analyses, versus reconstructed energy, with and without including the condition that the signal-to-background ratio has to be at least 5\%. The MAGIC reference is taken from \cite{ALEKSIC201676}}
    \label{fig:diff_sensitivity}
\end{figure}

 The result, for source-dependent and source-independent analysis, is shown in Fig. \ref{fig:diff_sensitivity}. The fluxes are given in fraction of the Crab Nebula flux (\enquote{Crab units}, C.U.). The shaded bands show the total uncertainty of the calculation, assuming a $\pm$1\% systematic uncertainty in the background normalization: the band edges are obtained by modifying the gamma-ray excess in each bin by $\pm(\sigma_\text{stat} + 0.01\;\text{N}_\text{off})$ where N$_\text{off}$ is the number of events after cuts in the off-source region.

 We also calculated the sensitivity using only the condition of 5-$\sigma$ statistical significance in 50 hours (re-optimizing the cuts), to show in which energy range (near threshold) the sensitivity is limited by the systematic uncertainty in background normalization. Note that in observations of e.g. pulsars in which the background can be estimated from the on-source events recorded in the off-pulse phase range (see section \ref{sec:pulsar}), we expect negligible background systematics, and hence gamma-ray excesses of well below 5\% of the background can be robustly detected.
 
The best integral sensitivity (for a Crab-like spectrum) in 50 hours is obtained for E $>$ 250 GeV, and is 1.1\% C.U. We also computed the integral sensitivity for a 0.5-hour exposure, given the fact that short-time observations of transient events are one of the main goals of LSTs, and we reach an integral sensitivity of 12.4\% C.U. above 250 GeV.

 Somewhat surprisingly, the performance of the source-independent and source-dependent analyses in terms of flux sensitivity is very similar, with the latter only slightly better at the highest energies. It seems that, with the current analysis, the introduction of the a-priori known direction of the point-like source in the event reconstruction does not significantly improve the background suppression capabilities.

 The MAGIC sensitivity \citep{ALEKSIC201676} is also shown for comparison in Fig. \ref{fig:diff_sensitivity}: as expected, despite the larger mirror area of \LST{} compared to that of the MAGIC telescopes, the advantages of stereoscopic reconstruction can be clearly seen in the plot. Above 100 GeV, MAGIC has a factor $\sim$1.5 better sensitivity on average. At lower energies the difference actually increases, despite the lower \LST{} threshold. The smallest difference is seen at the highest energies, a result of the much larger field of view of \LST{}, which provides larger reach in impact parameter.

\subsection{Crab Nebula Spectrum and Light Curve}
Aside from the metrics presented in previous sections, we assess the performance of the telescope by extracting the spectral energy distribution (SED) and light curve of the gamma-ray emission from the Crab Nebula, known to be stable in the VHE band, and comparing them with previous measurements reported by other instruments. DL3 data, containing gamma-like event candidates and the IRFs, are further processed using Gammapy v0.20 \citep{gammapy} to produce these high-level results for the two analysis approaches mentioned above. The \LST{} SEDs include also a small contribution from the pulsar at the lowest energies (estimated from \cite{MAGIC_Crab_TeV} to be $\simeq 10\%$ and 2\% of the total flux at 30 and 100 GeV respectively), which is smaller than the total uncertainty and has {\it not} been subtracted in this analysis. 

The event selection starts with an image \textit{intensity} cut of > 80 p.e. for the analysis of the entire dataset (relaxed to > 50 p.e. for the separate analysis of the post-August 2021 subset), as explained in section \ref{sec:data_sample}. The gamma-ray candidates are then chosen by applying energy-dependent \textit{gammaness} and angular cuts that keep a given percentage of the MC gamma-ray events in each bin of reconstructed energy. As baseline settings we decided to use 70$\%$ efficiency for both the \textit{gammaness} and the $\theta$ cuts. Besides, we set maximum values 0.95 for the \textit{gammaness} cut and $0.32 \degree$ for the $\theta$ cut. 
 
Since the dataset fully consists of observations performed in wobble mode, we can estimate the residual background in the signal region by using the event count in a control off-source sky region within the field of view, as explained in section \ref{sec:data_mc_comparison}.

\begin{figure*}
    \centering
    \includegraphics[width=0.49\textwidth]{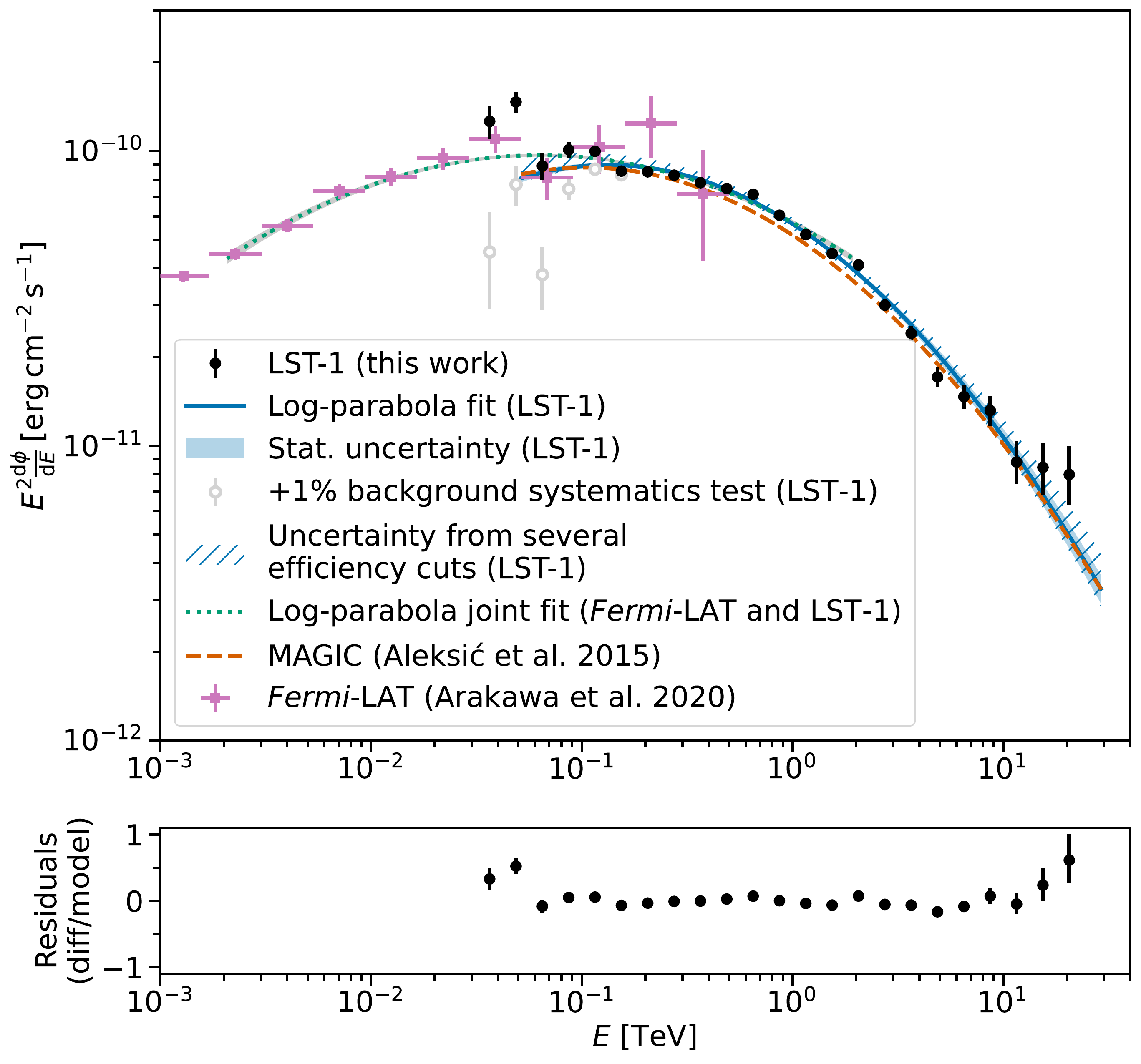}
    \includegraphics[width=0.49\textwidth]{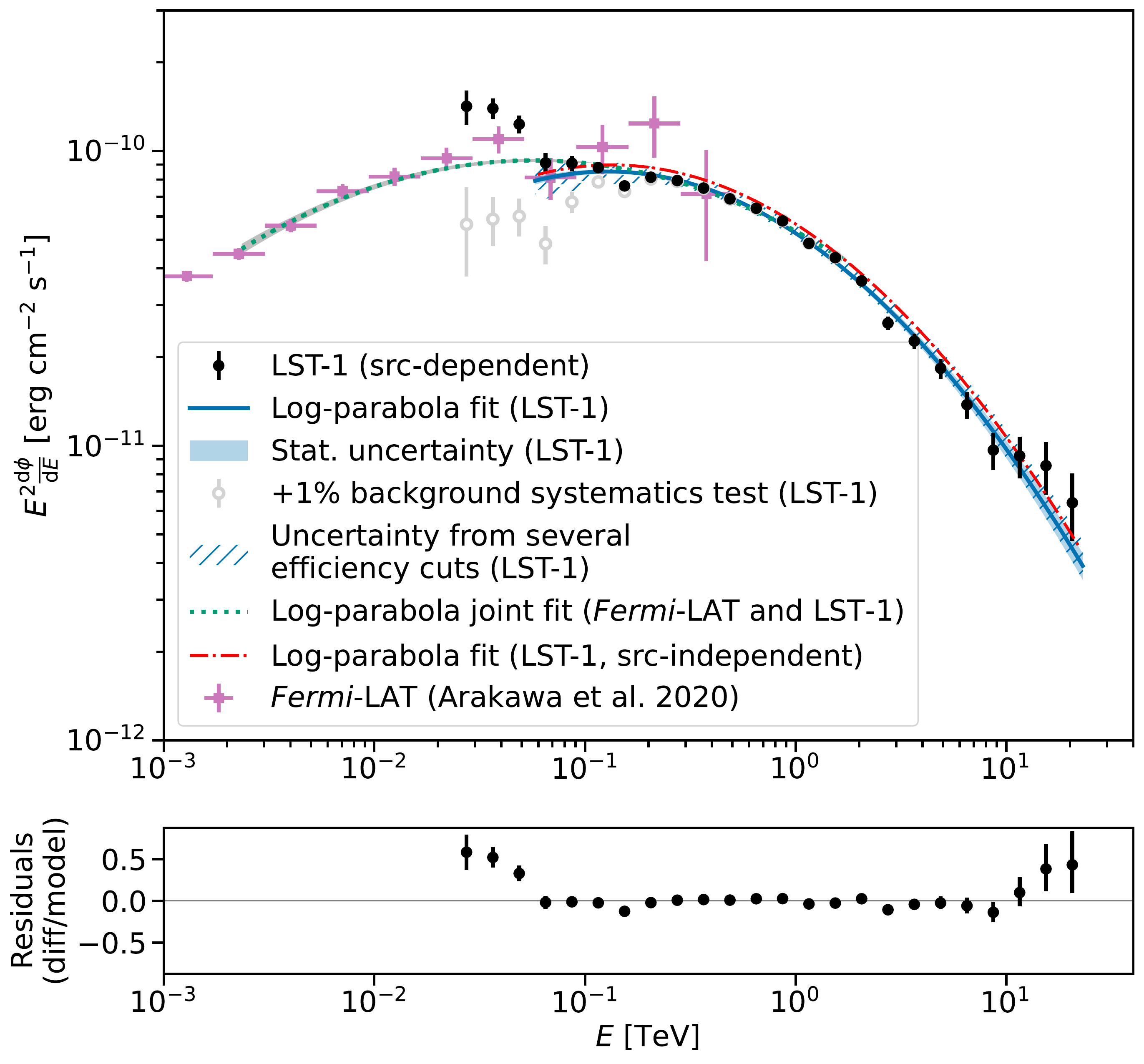}
    \caption{SED of the Crab Nebula for the entire \LST{} dataset obtained with the source-independent analysis (left panel) and source-dependent analysis (right panel). Flux points (black circles) and best-fit model (solid blue line) correspond to the dataset with a cut in image \textit{intensity} > 80 p.e., and energy-dependent \textit{gammaness} and $\theta/alpha$ (source-independent/source-dependent) selection cuts with 70\% gamma-ray efficiency. The solid error band illustrates the statistical uncertainty of the fit. Open markers represent the effect of increasing the background normalization by 1\%, to show that even such a small systematic error can have a large effect, well beyond the statistical uncertainty, on the flux at the lowest energies. The joint fit of the \textit{Fermi}-LAT and \LST{} spectra is represented by the dotted line (accompanied by its statistical uncertainty band). The SED model obtained with the source-independent analysis is also shown for comparison in the right panel (red dot-dash line).}
    \label{fig:crab_spectrum}
\end{figure*}

We then perform a forward-folding likelihood fit in the energy range 50 GeV - 30 TeV for straightforward comparison with the MAGIC reference \citep{CrabMAGIC} assuming a log-parabola spectral shape for the differential energy spectrum: 

\begin{equation} \label{eq:logpar}
    \dv*{\phi}{E}=f_{0} \cdot (E/E_{0})^{-\alpha-\beta \cdot \log(E/E_{0})} \,[{\rm cm}^{-2}\,{\rm s}^{-1} \,{\rm TeV}^{-1}] \rm{,}
\end{equation}

\noindent where $E_{0}=400$~GeV was chosen close to the decorrelation energy (energy at which the normalization of the spectrum, $f_0$, is least correlated with the other spectral parameters), and log is the natural logarithm. The best-fit model for the entire dataset is shown in the left panel of Fig.~\ref{fig:crab_spectrum}, and the resulting spectral parameters are listed in Table~\ref{tab:spectral_parameters}. Besides, using the Gammapy utility \texttt{FluxPointsEstimator}, we display the flux points calculated based on this spectral model considering eight bins per decade logarithmically spaced. The procedure followed to obtain the flux normalization in each energy bin is described in Sect. 3.5 of \citet{Acero2015}. Since the fitting range of the model starts at 50~GeV, lower-energy flux points up to 1~TeV are instead computed taking into account the joint model fit of the \textit{Fermi}-LAT and \LST{} datasets (see description below).

\begin{figure}
    \centering
    \includegraphics[width=0.49\textwidth]{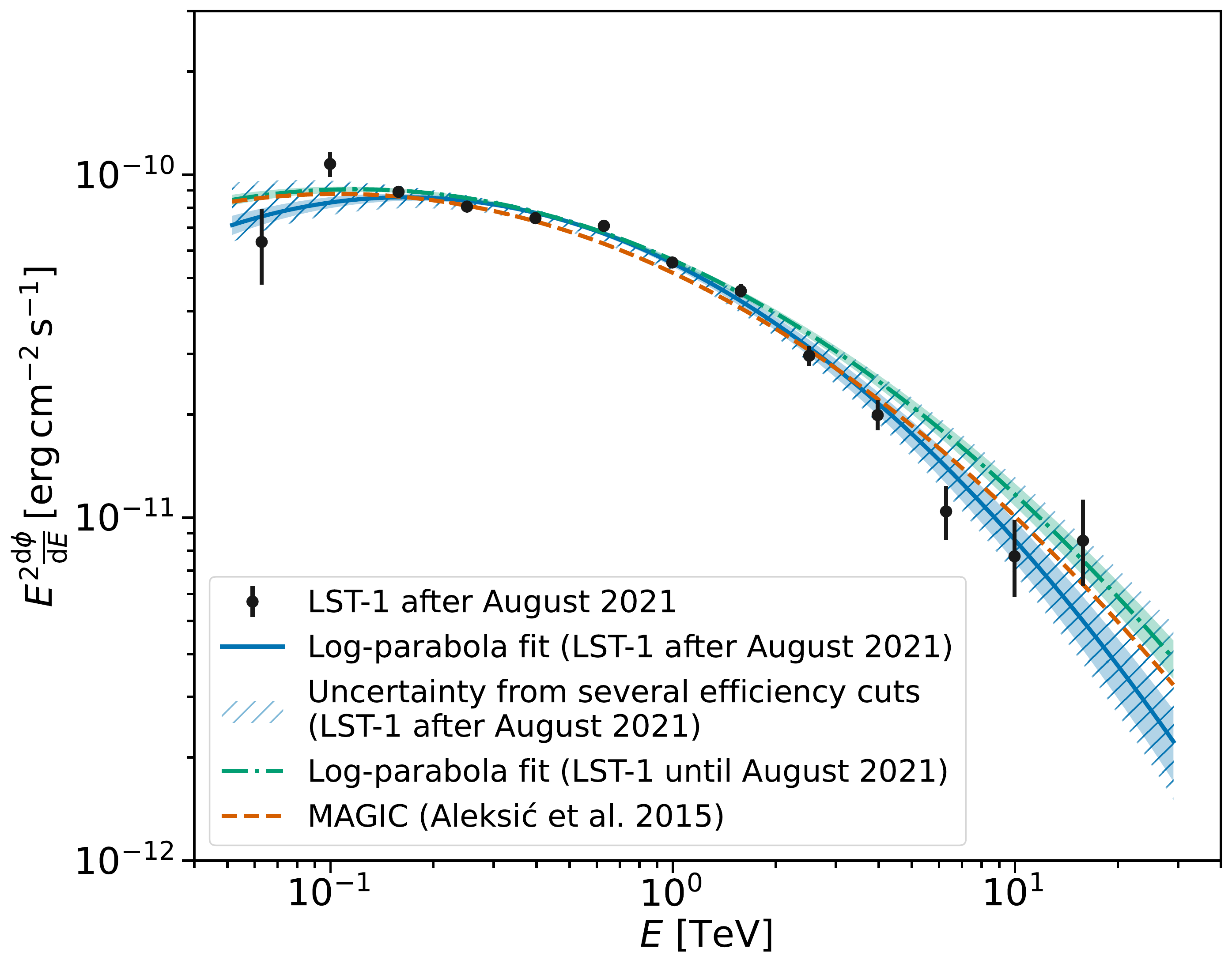}
    \caption{Comparison of Crab Nebula SED before and after the update of the \LST{} trigger settings in August 2021, with 70\%-efficiency \textit{gammaness} and $\theta$ selection cuts. The image \textit{intensity} cut is > 80~p.e. before August 2021, and > 50~p.e. afterwards (see text). Spectral points are shown for the post-August sample only. The best fit to a log-parabola model and the corresponding statistical error bands for both data samples are also displayed. The hatched region represents the effect of varying the efficiency of the signal selection cuts for the data sample taken after August 2021.}
    \label{fig:SED_comparison_prepost_Aug21}
\end{figure}

\begin{table*}
    \centering
    \begin{tabular}{c|c|c|c|c|c}
Analysis type & $f_0$ [TeV$^{-1}$cm$^{-2}$ s$^{-1}$] & $E_0$ [GeV]  & $\alpha$ & $\beta$ & $\chi^2$ / $N_{dof}$ \\
\hline
\hline
Source-independent & $(3.05 \pm 0.02) \times 10^{-10}$ & 400 & $2.25 \pm 0.01$ & $0.114 \pm 0.006 $ & 48.5 / 18\\ 
Source-dependent   & $(2.87 \pm 0.02) \times 10^{-10}$ & 400 & $2.26 \pm 0.01$ & $0.115 \pm 0.006 $ & 32.9 / 18\\ 
    \end{tabular}
    \caption{
    Spectral parameters for the source-independent and source-dependent analyses of the Crab Nebula in the energy range 50 GeV - 30 TeV assuming a log-parabolic parametrization.
    }
    \label{tab:spectral_parameters}
\end{table*}

In order to check that the SED model does not significantly change with the applied gamma-ray efficiencies, we obtained the SED for different combinations of \textit{gammaness} and $\theta$ selection cuts, with (40, 70, 90)\% gamma-efficiency for \textit{gammaness} and (70, 90)\% in the case of $\theta$. Tighter $\theta$ cuts are not advisable, given the discrepancies shown in Fig. \ref{fig:MC_vs_data_theta2}. The envelope of the resulting SEDs is shown as the hatched area in Fig. \ref{fig:crab_spectrum}. This area provides us with a rough estimate of the systematic uncertainty we may have from mismatches between the actual telescope performance and the MC simulation (in case of a significant mismatch, tighter signal selection cuts always result in underestimated fluxes).

In view of the behaviour of the low-energy spectral points, which lie significantly above the best-fit SED, we also evaluated the effect of a possible systematic error in the background estimation. The baseline assumption in the analysis is that the signal and the control regions have identical acceptance, and hence the event count in the latter is an unbiased estimate of the number of background events in the former. The open markers in Fig. \ref{fig:crab_spectrum} show how the spectral points would change when the background estimate is increased by 1\%. Such small increase in the background, which only affects the lowest-energy part of the SED, is enough to bring the anomalous spectral points well below the best-fit SED and the \textit{Fermi}-LAT measurements in the same energy range. This test highlights the limited background suppression capabilities of a single IACT near its threshold, resulting in gamma-ray excesses of only a few percent of the residual background, even for a source as bright as the Crab Nebula. Note that the large effect in the SED below 100 GeV of the 1\% background modification actually hints at a smaller background systematic error, if we take the \textit{Fermi}-LAT points as a reference. As a further check, we also compared the background rate after cuts in two off-source regions at the same distance from the center of the FoV (0.4$^\circ$), and equidistant from the Crab, and we obtain a difference between them of $0.5 \pm 0.2 \%$ (5557 $\pm$ 1534 events, for an average background of $1.18 \times 10^6$ events) for E$_{reco} <$ 63 GeV. 
We cannot claim this to be the systematic uncertainty of the background estimation at low energies, valid for all \LST{} observations, since this is expected to depend on the camera response homogeneity, and hence on details like the brightness and distribution of stars in the field of view.

The obtained \LST{} Crab spectrum is very close (within 10\% in flux) to the MAGIC reference \citep{CrabMAGIC}. Considering the systematic uncertainties in both measurements, no significant discrepancies are seen. The spectrum also connects smoothly with the most energetic part of the \textit{Fermi}-LAT spectrum \citep{Arakawa_2020}, as illustrated by the joint fit of the \textit{Fermi}-LAT and \LST{} SEDs to a log-parabola model shown in Fig.~\ref{fig:crab_spectrum}. The observed bump in the SED 
is commonly explained by the inverse Compton emission. From the joint fit, we estimate the position of the peak to be around 60 GeV. 
We note that the resulting statistics from this fit would not be meaningful because only statistical uncertainties are considered, whereas the systematic uncertainties, very relevant especially near the threshold, are not taken into account.

Additionally, we evaluated the SEDs obtained with \LST{} before and after setting up the current trigger settings in August 2021, which resulted in a lower and more stable energy threshold, as explained in section~\ref{sec:data_sample}. We split the dataset into two samples corresponding to the observations carried out before and after August 2021. The livetime of the data sample until and after August 2021 is 25.4~h and 8.8~h, respectively. Both SEDs are compared in  Fig.~\ref{fig:SED_comparison_prepost_Aug21}. As there are less data after August 2021, we use five bins per decade to calculate the flux points. We note that the post-August 2021 sample has a slightly lower flux, perhaps related to the lower average light collection efficiency estimated from muons (see section~\ref{sec:data_mc_comparison}). Moreover, this subset could be affected more significantly by external factors like non-optimal atmospheric conditions since its time interval is shorter. The agreement is better when soft cuts are applied (see the upper edge of the hashed blue band on Fig.~\ref{fig:SED_comparison_prepost_Aug21}), which could point to a slightly worse agreement of data and MC for this data subset.

We remark that the bulk of the post-August 2021 sample was recorded after the interruption of \LST{} operations due to the eruption of the Cumbre Vieja volcano in La Palma between September 19$^{\rm th}$ and December 13$^{\rm th}$ 2021.

\subsubsection{Spectrum Obtained with the Source-dependent Analysis}
We also computed the spectrum of the Crab Nebula using the source-dependent analysis to validate this analysis method. For this analysis, energy-dependent $\alpha$ cuts are applied to obtain a given gamma-ray efficiency, instead of the $\theta$ cut used in source-independent analysis. The right panel of Fig.~\ref{fig:crab_spectrum} shows the Crab Nebula SED obtained with source-dependent analysis for the whole dataset, and a cut in \textit{intensity} > 80 p.e. The flux points and best-fit models are computed with 70\% efficiency cuts for both \textit{gammaness} and $\alpha$. The best-fit model parameters are also indicated in Table~\ref{tab:spectral_parameters}. Although the flux of the best-fit model is statistically incompatible with that derived from the source-independent analysis, we must keep in mind that the uncertainties quoted in the table are just statistical. As in the source-independent analysis, event selection cuts are changed to obtain (40, 70, 90)\% gamma-efficiency for \textit{gammaness}, and (70, 90)\% in the case of $\alpha$, as an estimate of the systematic uncertainty related to data-MC discrepancies. The result is represented as the hatched band in Fig.~\ref{fig:crab_spectrum}. When these bands are considered, the SEDs from the source-dependent and the source-independent analyses are compatible.

 Also in the case of source-dependent analysis the lowest energy spectral points deviate significantly from the best-fit SED. But again, a sub-percent variation of the background normalization would be enough to make them match the fit. Once more, we emphasize the importance of considering potential background systematic uncertainties in the spectral analysis of IACT data (eventually, this should be included in the likelihood maximization as a nuisance parameter). Note that this is a potential issue not only for single IACTs - even a stereoscopic system, with much stronger background suppression, can face a similar problem with dimmer sources in long-term observations.

\subsubsection{Light Curve} 
\begin{figure*}
    \centering
    \includegraphics[width=0.487\textwidth]{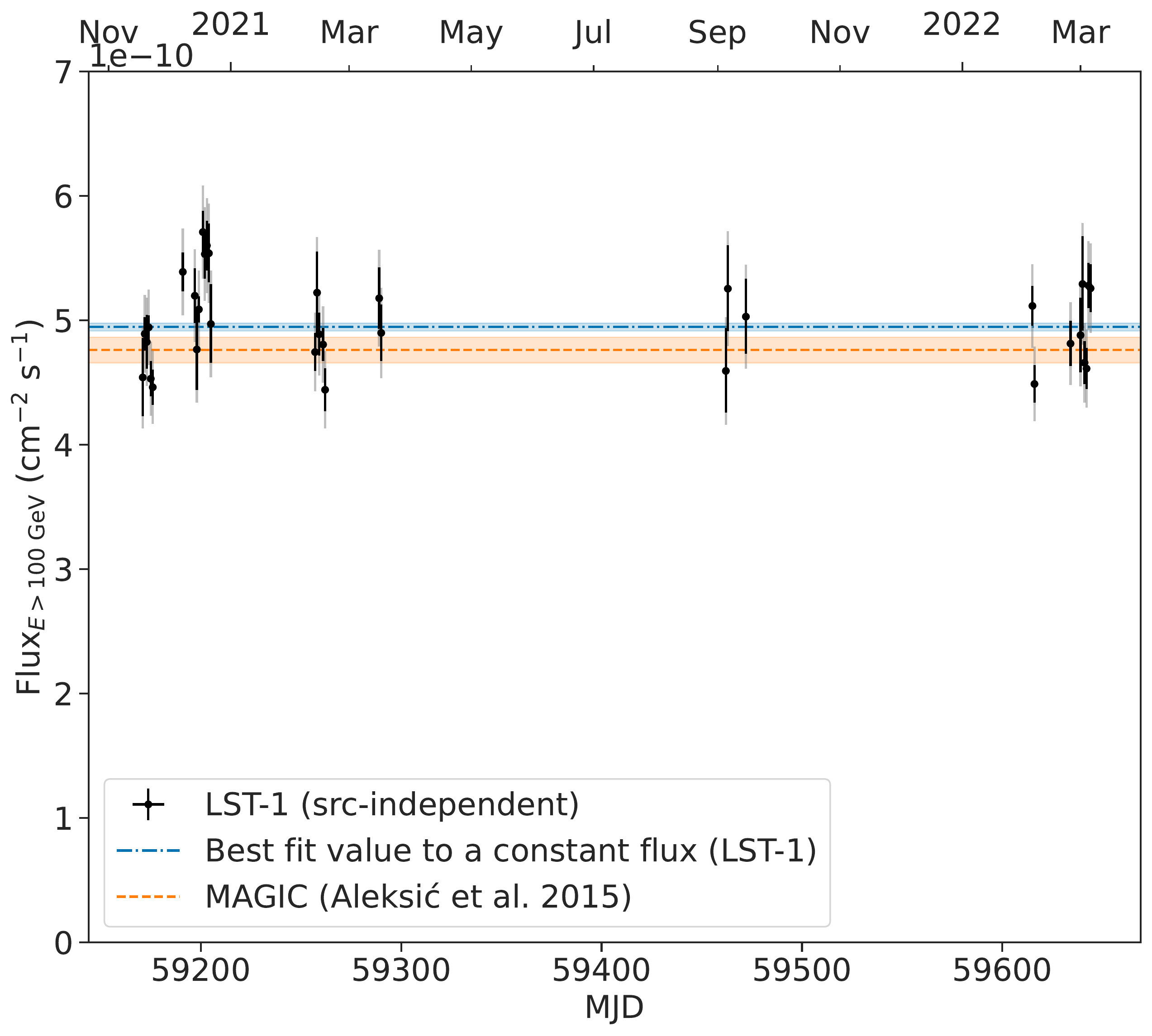}
    \includegraphics[width=0.497\textwidth]{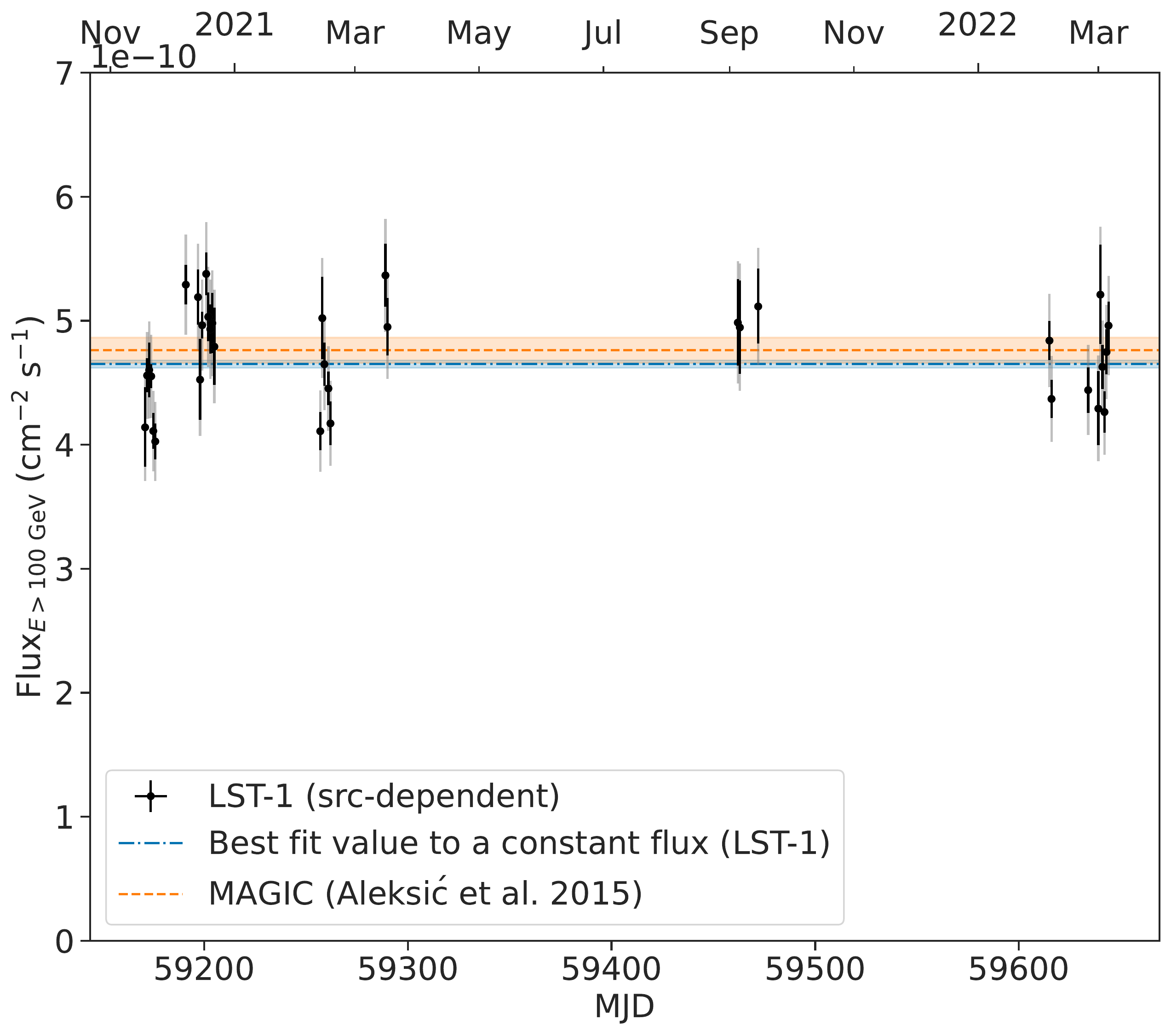}
    \caption{Crab Nebula light curve with 1-day bins above 100 GeV for source-independent (left) and source-dependent analyses (right). The dash-dotted line is the best fit to a constant flux. We also indicate the integral flux in the same energy range calculated from the log-parabola model reported in \cite{CrabMAGIC} with a dashed line. The black error bars correspond to the statistical errors. The gray ones include the systematic uncertainties (added in quadratic sum) assuming they are 6\% and 7\% of the flux values for source-independent and source-dependent analysis, respectively.}
    \label{fig:crab_lc}
\end{figure*}

In order to check the stability of the VHE gamma-ray flux from the Crab Nebula throughout the observations reported in this work ($\simeq$1.5 years), we calculated the daily light curve above 100 GeV, that is a safe minimum energy to avoid threshold effects, and displayed it in Fig.~\ref{fig:crab_lc} for both analysis approaches. We assume a log-parabola spectral model with the corresponding best-fit parameters indicated in Table~\ref{tab:spectral_parameters}. Flux points are fitted to a constant value of 
\begin{itemize}
\item[]{$F_{>100 \rm{~GeV}} = (4.95 \pm 0.03) \times 10^{-10} \rm{~cm}^{-2} \rm{~s}^{-1}$}
\item[]{$\;\;\;\;$with $\chi^2 / N_{dof} = 119.2 / 33$ (P-value=$1 \times 10^{-11}$)} 
\end{itemize}
for the source-independent analysis, and 
\begin{itemize}
\item[]{$F_{>100 \rm{~GeV}} = (4.65 \pm 0.03) \times 10^{-10} \rm{~cm}^{-2} \rm{~s}^{-1}$}
\item[]{$\;\;\;\;$with $\chi^2 / N_{dof} = 147.6 / 33$ (P-value=$2 \times 10^{-16}$)}
\end{itemize}
for the source-dependent analysis. Both results are, at face value, strongly incompatible with the (presumably) steady VHE flux of the nebula. However, only statistical uncertainties are considered - and from the tests performed with spectra (varying cut efficiencies and background normalization), it is clear that the total uncertainty must be significantly larger. 

In order to obtain a light curve "fully compatible" with a steady flux (P-value $\simeq 0.5$) we have to assume, for the source-independent analysis, an additional systematic uncertainty of 6\% on the nightly flux values, added in quadrature to the statistical uncertainty (see Fig.~\ref{fig:crab_lc}). In the case of the source-dependent analysis the value is 7\%. This level of systematics seems plausible, considering that no run-wise or night-wise IRFs (to account for variable observation conditions or telescope performance) have been used in the calculations. Obviously, these estimates, computed under the assumption that the Crab Nebula flux is constant at these energies, do not tell us anything about a possible overall systematic error affecting all nights in the sample.

\subsection{Crab Pulsar Phaseogram \label{sec:pulsar}} 
\begin{figure*}
\begin{center}
\includegraphics[width=0.98\textwidth]{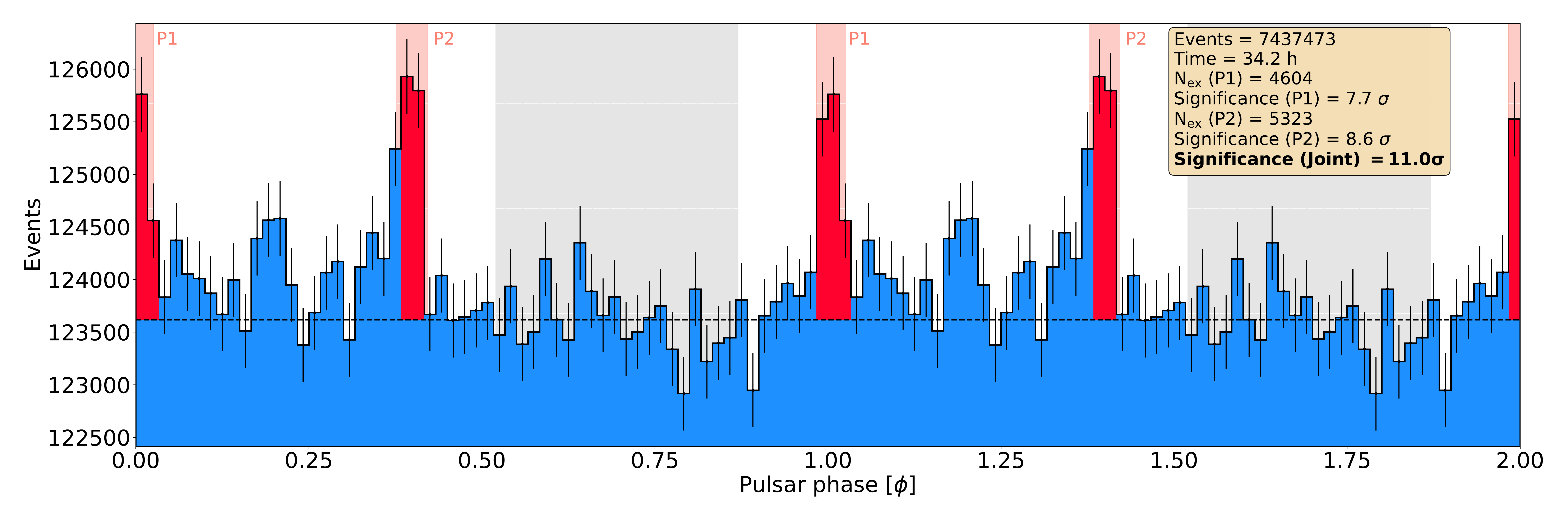}
\vspace{-0.3cm}
\caption{Crab pulsar phaseogram. The Crab pulsar phases definition is taken from \cite{magic_crab_2012}. We use the source-dependent analysis with fixed cuts for the whole energy range of $gammaness\ >\ 0.6$ and $\alpha\ <\ $12$^\circ$ and $intensity\ >\ 50$ p.e.}
\label{fig:crab_pulsar_phaseogram}
\end{center}
\end{figure*}

The observations of the Crab Nebula have as by-product another low-energy source that can be used to study its performance. The Crab pulsar (PSR J0534+220) is a young neutron star with a rotational period of 33 ms created after the supernova explosion SN1054. It has the second highest spin-down power known ($\dot{E}=$4.6 $\times$ 10$^{38}$ erg s$^{-1}$). It was first detected at VHE gamma rays by MAGIC \citep{MAGIC_Crab_25GeV} and over the years its spectrum was extended up to TeV energies \citep{Crab_VERITAS, magic_crab_2012, MAGIC_Crab_TeV}. The dataset used to search for pulsations is the same as in the rest of this article.  The Crab pulsar phases definition is taken from \cite{magic_crab_2012}. Both P1 and P2 peaks are significantly detected as it can be seen in Fig. \ref{fig:crab_pulsar_phaseogram}, produced with the source-dependent analysis. The calculation of the pulsar spectrum will require a more detailed treatment of the runs with non-standard trigger threshold settings, and is a work in progress that will be the subject of a future publication.

\section{Summary and Conclusions}
\label{sec:conclusions}
We presented in this paper the observations of the Crab Nebula with the CTA~\LST{} telescope performed during its commissioning period, and used them to evaluate the instrument performance in single-telescope mode.

The optical efficiency of the system, as determined with muon rings, is stable within $\pm5\%$ in the $\simeq$1.5 year span of the dataset shown in this paper. The trigger threshold, on the other hand, was not fully stable through this period, and is on average a little higher than its design value (reached only in August 2021). The trigger threshold for the current configuration is 20 GeV, which increases to $\simeq 30$ GeV after analysis cuts.

The standard source-independent analysis can reach an angular resolution better than $0.12^\circ$ for $E>1$ TeV using hard cuts (low efficiency). For the baseline cuts (70\% efficiency) used to derive the spectra and light curves presented in this paper, the angular resolution is $\simeq0.17^\circ$ for $E>1$ TeV and $\simeq 0.34^\circ$ at $E = 100$ GeV. The energy resolution reaches the level of 20\% for $E>$ 1 TeV, and 35\% at $E = 100$ GeV for low-zenith observations. Below 50 GeV, the source-dependent analysis provides slightly better energy resolution and smaller bias. 

In terms of flux sensitivity, the source-independent and source-dependent analyses provide similar performance, with the latter being slightly better at the highest energies. The 50-hour sensitivity above 100 GeV is about $\sim$1.5 times worse than that of the MAGIC telescopes that operate in a similar energy range in stereoscopic mode. The advantages of stereoscopic reconstruction are, as expected, not overcome by the larger light collection efficiency of the \LST{} (from its larger dish and more efficient camera). The optimal 50-h differential sensitivity achieved is $\sim$1.7 \% C.U. at 800 GeV using source-independent analysis. The best integral sensitivity of the instrument is 1.1\% C.U. above 250 GeV in 50 hours (12.4\% C.U. in 0.5 hours). 

We find that the Crab Nebula spectra derived using source-dependent and source-independent analysis are both very close (within 10\%) to the spectrum determined with the MAGIC telescopes. The fact that the spectral fit results (obtained using only statistical uncertainties) from the two methods are not statistically compatible clearly indicates that for a bright source like the Crab Nebula statistical uncertainties are sub-dominant with respect to systematics uncertainties, like those resulting from small data-MC discrepancies, or from inhomogeneities of the telescope response across the field of view, which may result in a biased background estimation.

We also find a smooth connection between the \LST{} spectrum and the $Fermi$-LAT one, especially when the possible systematics in the background normalization in the \LST{} near-threshold analysis are taken into account. The Crab Nebula night-wise light curves above 100 GeV derived using the two analyses are compatible with a steady emission if an additional systematic uncertainty in the flux values of $\simeq 7\%$ is assumed. Under the assumption that the VHE emission from the Crab Nebula was stable during the \LST{} observations, this again shows that systematic uncertainties in the calculated fluxes are non-negligible in an observation like the one presented here, and should become the focus of future analysis improvements.

\bigskip\bigskip
{\bf Acknowledgements:}
\par
The Camera time-stamping hardware was developed and provided by the Astroparticle and Cosmology laboratory (APC at Université Paris Cité, CNRS-IN2P3). 
\par

We gratefully acknowledge financial support from the following agencies and organisations:

Conselho Nacional de Desenvolvimento Cient\'{\i}fico e Tecnol\'{o}gico (CNPq), Funda\c{c}\~{a}o de Amparo \`{a} Pesquisa do Estado do Rio de Janeiro (FAPERJ), Funda\c{c}\~{a}o de Amparo \`{a} Pesquisa do Estado de S\~{a}o Paulo (FAPESP), Funda\c{c}\~{a}o de Apoio \`{a} Ci\^encia, Tecnologia e Inova\c{c}\~{a}o do Paran\'a - Funda\c{c}\~{a}o Arauc\'aria, Ministry of Science, Technology, Innovations and Communications (MCTIC), Brasil;
Ministry of Education and Science, National RI Roadmap Project DO1-153/28.08.2018, Bulgaria;
Croatian Science Foundation, Rudjer Boskovic Institute, University of Osijek, University of Rijeka, University of Split, Faculty of Electrical Engineering, Mechanical Engineering and Naval Architecture, University of Zagreb, Faculty of Electrical Engineering and Computing, Croatia;
Ministry of Education, Youth and Sports, MEYS  LM2015046, LM2018105, LTT17006, EU/MEYS CZ.02.1.01/0.0/0.0/16\_013/0001403, CZ.02.1.01/0.0/0.0/18\_046/0016007 and CZ.02.1.01/0.0/0.0/16\_019/0000754, Czech Republic; 
CNRS-IN2P3, the French Programme d’investissements d’avenir and the Enigmass Labex, 
This work has been done thanks to the facilities offered by the Univ. Savoie Mont Blanc - CNRS/IN2P3 MUST computing center, France;
Max Planck Society, German Bundesministerium f{\"u}r Bildung und Forschung (Verbundforschung / ErUM), Deutsche Forschungsgemeinschaft (SFBs 876 and 1491), Germany;
Istituto Nazionale di Astrofisica (INAF), Istituto Nazionale di Fisica Nucleare (INFN), Italian Ministry for University and Research (MUR);
ICRR, University of Tokyo, JSPS, MEXT, Japan;
JST SPRING - JPMJSP2108;
Narodowe Centrum Nauki, grant number 2019/34/E/ST9/00224, Poland;
The Spanish groups acknowledge the Spanish Ministry of Science and Innovation and the Spanish Research State Agency (AEI) through the government budget lines PGE2021/28.06.000X.411.01, PGE2022/28.06.000X.411.01 and PGE2022/28.06.000X.711.04, and grants PID2022-139117NB-C44, PID2019-104114RB-C31,  PID2019-107847RB-C44, PID2019-104114RB-C32, PID2019-105510GB-C31, PID2019-104114RB-C33, PID2019-107847RB-C41, PID2019-107847RB-C43, PID2019-107847RB-C42, PID2019-107988GB-C22, PID2021-124581OB-I00, PID2021-125331NB-I00; the ``Centro de Excelencia Severo Ochoa" program through grants no. CEX2019-000920-S, CEX2020-001007-S, CEX2021-001131-S; the ``Unidad de Excelencia Mar\'ia de Maeztu" program through grants no. CEX2019-000918-M, CEX2020-001058-M; the ``Ram\'on y Cajal" program through grants RYC2021-032552-I, RYC2021-032991-I, RYC2020-028639-I and RYC-2017-22665; the ``Juan de la Cierva-Incorporaci\'on" program through grants no. IJC2018-037195-I, IJC2019-040315-I. They also acknowledge the ``Atracción de Talento" program of Comunidad de Madrid through grant no. 2019-T2/TIC-12900; the project ``Tecnologi\'as avanzadas para la exploracio\'n del universo y sus componentes" (PR47/21 TAU), funded by Comunidad de Madrid, by the Recovery, Transformation and Resilience Plan from the Spanish State, and by NextGenerationEU from the European Union through the Recovery and Resilience Facility; the La Caixa Banking Foundation, grant no. LCF/BQ/PI21/11830030; the ``Programa Operativo" FEDER 2014-2020, Consejer\'ia de Econom\'ia y Conocimiento de la Junta de Andaluc\'ia (Ref. 1257737), PAIDI 2020 (Ref. P18-FR-1580) and Universidad de Ja\'en; ``Programa Operativo de Crecimiento Inteligente" FEDER 2014-2020 (Ref.~ESFRI-2017-IAC-12), Ministerio de Ciencia e Innovaci\'on, 15\% co-financed by Consejer\'ia de Econom\'ia, Industria, Comercio y Conocimiento del Gobierno de Canarias; the ``CERCA" program and the grant 2021SGR00426, both funded by the Generalitat de Catalunya; and the European Union's ``Horizon 2020" GA:824064 and NextGenerationEU (PRTR-C17.I1).
State Secretariat for Education, Research and Innovation (SERI) and Swiss National Science Foundation (SNSF), Switzerland;
The research leading to these results has received funding from the European Union's Seventh Framework Programme (FP7/2007-2013) under grant agreements No~262053 and No~317446;
This project is receiving funding from the European Union's Horizon 2020 research and innovation programs under agreement No~676134;
ESCAPE - The European Science Cluster of Astronomy \& Particle Physics ESFRI Research Infrastructures has received funding from the European Union’s Horizon 2020 research and innovation programme under Grant Agreement no. 824064.

\software{The analysis and figures of this manuscript were produced using the following open-access software tools: Astropy \citep{astropy2013, astropy2018}, Matplotlib \citep{matplotlib}, NumPy \citep{numpy}, Scikit-learn \citep{scikit-learn} and SciPy \citep{scipy}.\newline
All numerical data of the figures in the paper, and scripts to reproduce them, are available at \cite{lopez_coto_ruben_2023_8159146}
}\newline
\newline
\newline
This paper has gone through internal review by the CTA Consortium.
\newline
\newline
{\bf Author Contribution}
\newline
R. L\'opez-Coto: project coordination, muon ring analysis, sensitivity estimation, Crab pulsar analysis. A. Moralejo: project coordination, data sample selection, data-MC cross-validation,  sensitivity estimation. D. Morcuende: data analysis, Crab Nebula spectrum and light curve (source-independent approach). S. Nozaki: data analysis, Crab Nebula spectrum and light curve (source-dependent approach). T. Vuillaume:  Monte Carlo processing, Random Forest generation, computation of the Instrument Response Functions. All five corresponding authors above have participated in the paper drafting and edition.
The rest of the authors have contributed in one or several of the following ways: design, construction, maintenance and operation of the instrument(s) used to acquire the data; preparation and/or evaluation of the observation proposals; data acquisition, processing, calibration and/or reduction; production of analysis tools and/or related Monte Carlo simulations; discussion and approval of the contents of the draft.


\newpage

\bibliographystyle{aasjournal}
\bibliography{mybibfile}

\appendix

\section*{Appendix}

\subsection*{Random forests features}

The parameters used to train the random forest models are listed below (see also Fig. \ref{fig:image:sketch}):

\begin{itemize}
    \item \textit{log\_intensity}: decimal logarithm of \textit{intensity}, the sum of the charges (in photo-electrons) of the pixels which survive the image cleaning. All other image parameters listed below are calculated using the same set of pixels.
    \item \textit{width, length}: width and length of the ellipsoid, i.e. the second-order moments of the charge distribution along the minor and major axes of the image. 
    \item \textit{wl}: width over length ratio
    \item \textit{x, y}: coordinates of the image centroid (first order moments) in the camera frame
    \item \textit{skewness}: skewness of the image
    \item \textit{kurtosis}: kurtosis of the image
    \item \textit{time\_gradient}: gradient of the signal arrival times, computed along the main axis of the image
    \item \textit{leakage\_intensity\_width\_2}: fraction of the total charge of the image which is recorded on pixels at the edge of the camera, or their next neighbors.
    \item\textit{az\_tel, alt\_tel}: telescope pointing direction (azimuth and altitude respectively) 
    \item \textit{disp\_norm}: distance between \textit{(x,y)} and the point along the major image axis which is closest to the true (nominal) source position.
    \item \textit{disp\_sign}: indicates on which side of the image centroid \textit{(x,y) the true source position lies.}
    \item \textit{dist}: distance between \textit{(x,y)} and the true (nominal) source position on the camera.
    \item \textit{reco\_disp\_norm\_diff}: abs$(dist - reco\_disp\_norm)$
    \item \textit{reco\_disp\_sign\_correctness}: estimated probability (obtained from the {\it disp\_sign} random forest using the \texttt{predict\_proba} method of \texttt{scikit-learn})
    that the known source position (on either side of the centroid along the major axis) is the correct one for the given image.
\end{itemize}

\subsection*{Random forests hyperparameters}

The random forests regressors (for the energy and disp reconstruction) are trained using \texttt{scikit-learn} v1.0 with the following hyperparameters:
\begin{itemize}
    \item \texttt{max\_depth}=30
    \item \texttt{min\_sample\_leaf}=10
    \item \texttt{n\_estimators}=150
    \item \texttt{max\_depth}=\texttt{true}
    \item \texttt{criterion}=\texttt{squared\_error}
    \item \texttt{max\_features}=\texttt{auto}
    \item \texttt{max\_leaf\_nodes}=\texttt{null}
    \item \texttt{min\_impurity\_decrease}=0.0
    \item \texttt{min\_samples\_split}= 10
    \item \texttt{min\_weight\_fraction\_leaf}=0.0.
\end{itemize}

The classifier uses the same parameters, with the following differences: \texttt{n\_estimators}=100, \texttt{criterion}=\texttt{gini}. Notably, the \texttt{max\_depth}, \texttt{n\_estimators}, \texttt{min\_sample\_leaf} and \texttt{min\_sample\_split} parameter values have been optimised using a grid search to minimize computing resources without compromising the models performances.



\end{document}